\documentclass[12pt,preprint]{aastex}
\usepackage{graphicx}

\defcitealias{assef09}{Paper~I}

\begin{document}

\title{The Mid-IR and X-ray Selected QSO Luminosity Function}

\author{R.J.~Assef\altaffilmark{1}, 
  C.S.~Kochanek\altaffilmark{1},
  M.L.N.~Ashby\altaffilmark{2},
  M.~Brodwin\altaffilmark{2,3},
  M.J.I.~Brown\altaffilmark{4},
  R.~Cool\altaffilmark{5},
  W.~Forman\altaffilmark{2},
  A.H.~Gonzalez\altaffilmark{6},
  R.C.~Hickox\altaffilmark{2,7},
  B.T.~Jannuzi\altaffilmark{8},
  C.~Jones\altaffilmark{2},
  E.~Le~Floc'h\altaffilmark{9},
  J.~Moustakas\altaffilmark{10},
  S.S.~Murray\altaffilmark{2},
  D.~Stern\altaffilmark{11}
}

\affil{
  \altaffiltext{1} {Department of Astronomy, The Ohio State
    University, 140 W.\ 18th Ave., Columbus, OH 43210
    [email:{\tt{rjassef@astronomy.ohio-state.edu}}]}
  \altaffiltext{2} {Harvard-Smithsonian Center for Astrophysics, 60
    Garden St., Cambridge, MA 02138}
  \altaffiltext{3} {W. M. Keck Postdoctoral Fellow at the
    Harvard-Smithsonian Center for Astrophysics}
  \altaffiltext{4}{School of Physics, Monash University, Clayton 3800,
    Victoria, Australia}
  \altaffiltext{5} {Peyton Hall, Princeton University, Princeton, NJ
    08540}
  \altaffiltext{6} {Department of Astronomy, Bryant Space Science
    Center, University of Florida, Gainesville, FL 32611}
  \altaffiltext{7} {Department of Physics, Durham University, Durham
    DH1 3LE, UK}
  \altaffiltext{8}{KPNO/NOAO, 950 N. Cherry Ave., P.O. Box 26732,
    Tucson, AZ 85726}
  \altaffiltext{9} {CEA-Saclay, Service d'Astrophysique, Orme des
    Merisiers, Bat.709, 91191 Gif-sur-Yvette, FRANCE}
  \altaffiltext{10} {Center for Astrophysics and Space Sciences
    University of California, San Diego 9500 Gilman Drive La Jolla,
    California, 92093-0424}
  \altaffiltext{11} {Jet Propulsion Laboratory, California Institute of
    Technology, 4800 Oak Grove Drive, Mail Stop 169-506, Pasadena, CA
    91109}
}

\begin{abstract}
  We present the $J-$band luminosity function of 1838 mid-infrared and
  X-ray selected AGNs in the redshift range $0 < z < 5.85$. These
  luminosity functions are constructed by combining the deep
  multi-wavelength broad-band observations from the UV to the mid-IR
  of the NDWFS Bo\"otes field with the X-ray observations of the
  XBo\"otes survey and the spectroscopic observations of the same
  field by AGES. Our sample is primarily composed of IRAC-selected
  AGNs, targeted using modifications of the \citet{stern05} criteria,
  complemented by MIPS 24$\mu$m and X-ray selected AGNs to alleviate
  the biases of IRAC mid-IR selection against $z\sim 4.5$ quasars and
  AGNs faint with respect to their hosts. This sample provides an
  accurate link between low and high redshift AGN luminosity functions
  and does not suffer from the usual incompleteness of optical samples
  at $z\sim 3$. We use a set of low resolution SED templates for AGNs
  and galaxies presented in a previous paper by \citet{assef09} to
  model the selection function of these sources and apply host and
  reddening corrections. We find that the space density of the
  brightest quasars strongly decreases from $z=3$ to $z=0$, while the
  space density of faint quasars is at least flat, and possibly
  increasing, over the same redshift range. At $z>3$ we observe a
  decrease in the space density of quasars of all brightnesses. We
  model the luminosity function by a double power-law and find that
  its evolution cannot be described by either pure luminosity or pure
  density evolution, but must be a combination of both. We
  used the bright end slope determined by \citet[][2QZ]{croom04} as a
  prior to fit the data in order to minimize the effects of our small
  survey area. The bright-end power-law index of our best-fit model
  remains consistent with the prior, while the best-fit faint-end
  index is consistent with the low redshift measurements based on the
  2QZ and 2SLAQ \citep{croom09} surveys. Our best-fit model generally
  agrees with the number of bright quasars predicted by other LFs at
  all redshifts. If we construct the QSO luminosity function using
  only the IRAC-selected AGNs, we find that the biases inherent to
  this selection method significantly modify the behavior of the
  characteristic density $\phi_*(z)$ only for $z<1$ and have no
  significant impact upon the characteristic magnitude $M_{*,J}(z)$.

\end{abstract}


\section{Introduction}\label{sec:intro}

Galaxies with active galactic nuclei (AGN) are among the brightest
objects in the universe. Their light is thought to come from material
accreting onto a super massive black hole (SMBH) that resides in the
center of their host galaxies, so characterizing the AGN population
and its evolution across cosmic time is a direct constraint on the
formation and evolution of SMBHs. Recent theoretical studies
\citep[e.g.,][]{hopkins05,dimatteo05,hopkins06,croton06,bower06} have
suggested AGN activity may play an integral role in the evolution of
galaxies, so understanding AGN evolution may also constrain the
paradigm of galaxy evolution. The simplest way to characterize these
objects is by studying their spectral properties, their luminosity
function, and how it evolves with redshift.

What we observe from an AGN depends on its luminosity relative to its
host and the degree to which it is obscured by dust
\citep[e.g.,][]{antonucci93}. Broad line AGN are relatively unobscured
and divided into quasars and Type I Seyferts depending on the
luminosity of the AGN. For the bright quasars it is far more
difficult to observe the host galaxy than in the fainter
Seyferts. Unification models posit that the narrow line Type II
Seyferts are the same physical objects viewed from a direction along
which the nucleus is obscured by dust. Absorption by dust reprocesses
the shorter wavelength radiation from the accretion disc into
mid/far-IR continuum radiation. Hard radiation escaping in other
directions produces the narrow, high ionization lines characteristic
of these objects. If the absorbing column becomes too thick, the
nucleus can only be seen either in hard X-rays or radio with all other
radiation reprocessed into the infrared.

AGN are relatively rare objects compared to galaxies and
stars. Combined with their broad range of spectral properties, it is
difficult to select large uniform samples across the broad redshift
ranges needed to accurately characterize their evolution with cosmic
time. Broad-band photometric redshifts for AGNs are not nearly as
accurate as those for galaxies \citep[e.g.,][]{rowan08,assef09}, so
the samples selected from broad-band surveys must also be observed
spectroscopically in order to confirm their nature and measure their
redshifts. Medium- and narrow-band surveys can yield accurate
photometric redshifts and circumvent the need for spectroscopic
confirmation \citep{wolf03,salvato09}, but these surveys are
uncommon. Most surveys used to derive QSO luminosity functions (QLFs)
select targets based on optical broad-band colors and morphology. Such
surveys work remarkably well at low and high redshifts but suffer from
high incompleteness at $2.5 \lesssim z \lesssim 3.5$ where the optical
colors of quasars are very similar to those of stars
\citep{fan99}. Several studies
\citep[e.g.,][]{richards06b,brown06,croom09} have shown that the
density of bright quasars peaks at $z\sim 2$, so not being able to
uniformly build the QLFs from redshifts well below to well above this
peak is a limitation for understanding the evolution of nuclear
activity with cosmic time. Surveys based on X-ray, mid-IR and radio do
not have these limitations, but they are generally either too broad
and shallow or narrow and deep to follow the full sweep of quasar
evolution with cosmic time.

The AGN and Galaxy Evolution Survey \citep[AGES;][]{kochanek09} is a
redshift survey in the NOAO Deep Wide-Field Survey
\citep[NDWFS;][]{ndwfs99} Bo\"otes field using the multi-object
spectrograph Hectospec \citep{fabricant05} at the MMT. Using the deep
multi-wavelength observations of the NDWFS Bo\"otes field to select
targets, AGES obtained spectra for $\sim 6000$ AGN candidates and
$\sim 20000$ galaxies with $I<22.5$. In this paper, we present the
rest-frame $J-$band luminosity function derived from the
multi-wavelength photometric observations and the AGES spectroscopic
observations of 1838 AGNs in the NDWFS Bo\"otes field selected based
on their Spitzer IRAC \citep{fazio04} mid-IR colors, their MIPS
\citep{rieke04} 24$\mu$m fluxes, their X-ray counts in the XBo\"otes
survey \citep{xbootes}, and their optical morphologies. These
observations allow us to accurately study the evolution of the QLF
from $z = 0$ to $z = 5.6$. In a companion paper
\citep[][\citetalias{assef09}]{assef09} we present a set of empirical
SED templates that range from 0.03 to 30$\mu$m for galaxies and
AGNs. We use these templates to model the AGES AGN selection function,
to calculate absolute magnitudes and to correct for contamination from
the host galaxy and extinction. In \citet{dai08}, we present the
mid-IR LFs of galaxies in the NDWFS Bo\"otes field, while
\citet{cool10} presents the optical LF of galaxies and
\citet{rujopakarn10} presents their 24$\mu$m LF.

During the course of this paper we will interchangeably use the terms
AGN and QSO to refer to all objects with active nuclei, regardless of
inclination and luminosity. The technical differences between a QSO
and a non-QSO AGN is a matter of contention in the current literature,
but our sample spans a luminosity range large enough to encompass both
groups, rendering a differentiation meaningless for our current
purposes. The paper is organized as follows: in \S\ref{sec:data} we
describe the photometric and spectroscopic observations we use; in
\S\ref{sec:QSO_selection} we detail the AGES AGN selection criteria
and study the completeness of the spectroscopic observations; and in
\S\ref{sec:qso_lfs} we construct the mid-IR and X-ray selected QSO
luminosity function across the redshift range of our sample, study its
evolution and compare our results to previous studies at low and high
redshifts. Two appendices outline the mathematics of including X-ray
selected sources in our LFs and expand on the comparison between our
results and other estimates of quasar LFs. We use an $\Omega_M=0.3$,
$\Omega_{\lambda}=0.7$ and $H_0=73\ (\rm km\ \rm s^{-1}\ \rm
Mpc^{-1})$ cosmology throughout the paper.

\section{Data}\label{sec:data}

The data set we use is described extensively in \citet{assef08} and in
\citetalias{assef09}. Here we give a short description of it and refer
the reader to those works for details. Our data consists of the
extensive multi-wavelength photometric and spectroscopic observations
of the NOAO Deep Wide-Field Survey \citep[NDWFS;][]{ndwfs99} Bo\"otes
field. The NDWFS survey observed this field in the $B_{W}$, $R$, $I$
and $K$ bands. To this photometry, we have added observations, from
blue to red, in the X-rays from the XBo\"otes survey \citep{xbootes},
in the FUV and NUV channels of GALEX \citep[\citealt{martin05};
release GR5,][]{morrissey07}, in the $z$-band from the zBo\"otes
survey \citep{cool07}, in $J$ and $K_s$ from the Flamingos
Extragalactic Survey \citep[FLAMEX;][]{flamex06}, in all four IRAC
channels ([3.6], [4.5], [5.8] and [8.0]) from the Spitzer Deep
Wide-Field Survey \citep[SDWFS;][]{ashby09}, and from the 24$\mu$m
channel of MIPS \citep{weedman06}. Throughout this paper we will keep
the conventions of each survey. All magnitudes are in the Vega system
except for the AB magnitudes used by GALEX (FUV and NUV) and zBo\"otes
($z$-band).

The AGN and Galaxy Evolution Survey \citep[AGES;][]{kochanek09}
obtained optical spectra for approximately 26000 objects in the NDWFS
Bo\"otes field, and of these, about 6000 were targeted as AGN
candidates. Objects were targeted based on a broad range of
photometric criteria. In particular, AGNs are divided into many
uniform (non-exclusive) samples, based on their X-ray counts, radio
emission, 24$\mu$m MIPS fluxes and IRAC colors \citepalias[for details
see][\citealt{kochanek09}]{assef09}. In this paper we focus on the
IRAC, MIPS and X-ray selected samples, and in the next section we
discuss the different criteria by which these objects were
targeted. We limit our sample to the main survey area and we eliminate
areas close to bright stars, leaving an effective survey area of 7.47
square degrees.

We model the SEDs of each object using the templates derived in
\citetalias{assef09}. Note that the X-ray fluxes are not used in the
SED fits. Unlike \citetalias{assef09}, we use photometry measured in
3\arcsec\ diameter apertures instead of 6\arcsec\ in order to match
the photometry used by AGES for spectroscopic targeting. We also drop
the requirement of a minimum number of bands with detections or upper
bounds (8 in \citetalias{assef09}) in order to not bias our
sample. Nonetheless, all objects in our catalog, as defined in the
next section, have at least 6 bands of observed photometry and at
least 1 upper limit, which is enough for a well constrained SED fit
\citepalias[see][]{assef09}.

\section{Sample Selection}\label{sec:QSO_selection}

AGES used several different criteria to target AGNs for spectroscopic
observations. In this Section we discuss in detail the selection of
the sample we use in \S\ref{sec:qso_lfs} to construct luminosity
functions. In general, the bulk of the objects we consider have been
selected by their IRAC colors, but because this selection has some
important biases at low and high redshift, we complement it with
objects targeted as AGN by their X-ray counts and by their MIPS
fluxes.

\subsection{IRAC Colors AGN Selection}\label{ssec:IRAC_QSO_selection}

\begin{figure}
  \begin{center}
    \plotone{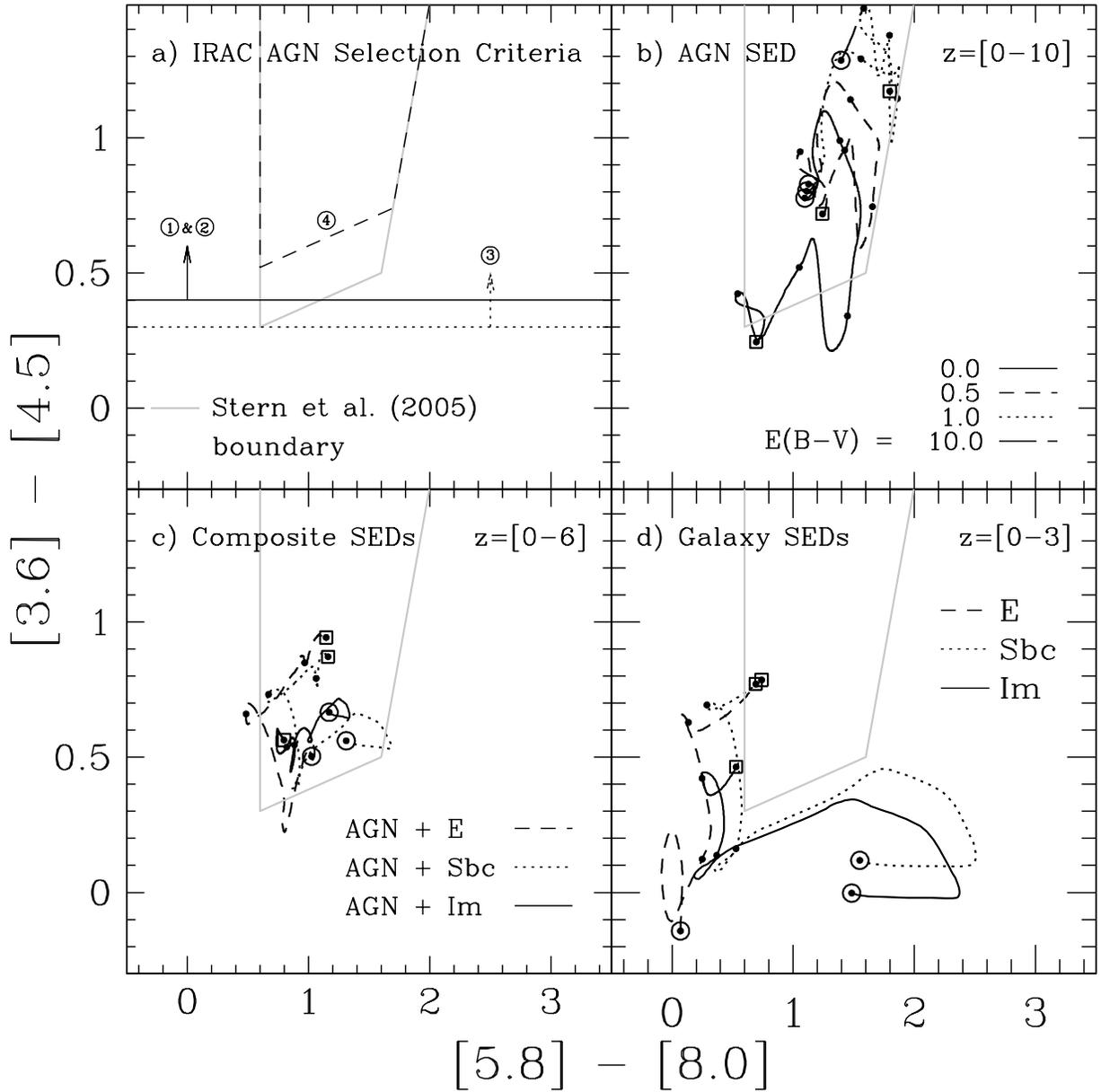}
    \caption{{\it{a)}} IRAC AGN selection criteria as detailed in
    \S\ref{sec:QSO_selection} (we do not show criterion 5 as we do not
    consider it for building the AGN luminosity functions). {\it{b)}}
    IRAC color-color track of a pure AGN (i.e. no galaxy host)
    according to our templates. The tracks are shown for the redshift
    range $z = 0 - 10$ and four different levels of extinction. For a
    single track, the bulls-eye marks $z=0$ and each successive black
    dot marks a redshift increase of 2. {\it{c)}} Same as b), but for
    objects whose SED is, bolometrically, half AGN and half galaxy for
    each galaxy template separately, and only in the redshift range
    0--6. For the AGN we have assumed no reddening. {\it{d)}} Same as
    b) and c) but for each galaxy template alone. Here the dots mark a
    redshift increase of 1 instead of 2, and we only show up to $z=3$,
    as galaxies at higher redshifts are not relevant to our study.}
    \label{fg:qso_tracks}    
  \end{center}
\end{figure}

\citet[][but also see \citealt{lacy04}]{stern05} empirically defined a
region in the IRAC color-color diagram that encompasses the locus of
QSOs and minimizes contamination by normal galaxies, although at the
expense of eliminating some active objects \citepalias[see
\citealt{gorjian08}, \citealt{donley08},][]{assef09}. Based on Spitzer
observations of the NDWFS field, AGES targeted prospective AGNs for
spectroscopy using a set of 5 IRAC criteria that slightly differed
from that of \citet{stern05} in order to increase or explore the
completeness of the sample. These criteria are shown in Figure
\ref{fg:qso_tracks} and can be summarized as:

\begin{enumerate}
  \item Optical point sources with observed magnitude $[3.6] \leq 18$
  and color $[3.6] - [4.5] \geq 0.4$.

  \item Optical point sources with observed magnitude $18 < [3.6] \leq 18.5$
  and color $[3.6] - [4.5] \geq 0.4$, with the additional constraint of
  either $I-[3.6] \geq 3$ or $z-[3.6] \geq 3$ in order to limit
  stellar contamination.

  \item Optical point sources with observed magnitude $[3.6] \leq 18$
  and color $0.3\leq [3.6] - [4.5] \leq 0.4$, again with the
  additional constraint of either $I-[3.6] \geq 3$ or $z-[3.6] \geq 3$
  to limit stellar contamination.

  \item Optically extended sources with observed magnitudes $I \leq
  20$ and $[3.6] \leq 18.5$, and IRAC colors that put them inside the
  region shown in Figure \ref{fg:qso_tracks}. This is a conservative
  version of the region defined by \citet{stern05} in which the lower
  boundary is shifted 0.1 mag redwards.

  \item Optically extended sources with observed magnitude $I \leq
  22.5$ that otherwise satisfy the same constraints as in 4.

\end{enumerate}

\noindent All colors are based on 3\arcsec\ aperture photometry
(corrected for the PSF) and all total magnitudes correspond to
{\tt{SExtractor}} \citep{sextractor96} ``auto'' magnitudes. Optical
point sources are defined as objects that have a {\tt{SExtractor}}
stellarity index greater than 0.8 in any of the optical bands, namely
$B_{W}$, $R$, $I$ and $z$. The separation between optically extended
and point sources should be quite reliable, as the 5$\sigma$ depth of
the depth of NDWFS is 25.5 magnitudes in $I-$band, 3 magnitudes
fainter than the $I = 22.5$ limit of AGES. 

The first criterion targets all point sources with a [3.6]--[4.5]
color that would roughly put them inside the region defined by
\citet{stern05} but disregards the information in the other two IRAC
bands. The lower sensitivity of the two longer wavelength bands means
that many more objects can be targeted based on the single
[3.6]--[4.5] color than by requiring both colors. For the magnitudes
we consider, we need not worry about the optically faint high redshift
galaxies lying to the left of the \citet{stern05} criterion in Figure
\ref{fg:qso_tracks} \citepalias[see \citealt{stern05},
][]{assef09}. Criterion 2 targets fainter point sources with an extra
constraint designed to eliminate normal stars as the color
uncertainties increase, while criterion 3 attempts to target AGNs lost
by the first two criteria due to emission lines passing through the
$[3.6]$ channel at high redshifts, most notably $\rm{H}\alpha$ at
$z\simeq 4.5$ \citepalias[see the discussion below, \S4.3 of][and
\citealt{richards09}]{assef09}.

Extended sources were targeted for spectroscopy if their IRAC colors
put them inside the region defined by the modified \citet{stern05}
boundaries of criteria 4 and 5. However, objects under criterion 5
were observed at a significantly lower priority than under criterion
4, and hence their spectroscopic completeness is very low. For this
reason, we will only consider objects selected based on criteria 1--4
in the rest of the paper. This means that extended sources are limited
to $I<20$. Similarly, point sources with $I>21.5$ were also observed
at a lower priority than their brighter counterparts, resulting in
significant incompleteness beyond this magnitude. To avoid dealing
with the effects of large incompleteness, we adopt a faint magnitude
limit of $I=21.5$ for our point source AGN catalog. Note that, by
doing so, we exclude the $z=6.12$ quasar found by \citet{stern07} from
our sample, as it has an $I$-band magnitude of $\approx 22$, but we
keep all other $z>5$ quasars found in the AGES sample \citep{cool06}.

Of the 1937 objects selected by these criteria, most correspond to
galaxies with active nuclei, but there is some contamination by
non-active sources. We have visually inspected all spectra and
eliminated all objects that did not show broad emission lines unless
they were classified as Type II AGNs by \citet{moustakas09} or were
detected in X-rays with 4 or more counts (see
\S\ref{sec:xray_selection}), reducing our sample to 1459 ``real''
AGNs. We consider for this purpose as Type II AGNs all objects that
have narrow line ratios above the limit of \citet{kauffmann03}, which
includes objects that can potentially be powered solely by star
formation \citep[see][]{kewley01}. Note that the Type II
classification of \citet{moustakas09} requires that H$\alpha$ is
contained in the wavelength range of the spectrum, effectively
limiting this classification to $z<0.4$. Of the 478 IRAC-selected
objects we have dropped, only 114 (24\%) are located inside the
original \citet{stern05} selection region, confirming that the
majority of these objects are contaminants due to the more relaxed
selection criteria of AGES compared to \citet{stern05}. The sources
inside the \citet{stern05} diagram are likely a combination of
contaminants due to photometric errors and ``real'' Type 2 AGN missed
by the BPT classification (see discussion in \S\ref{ssec:full_sample}
about the biases this may introduce in our results).

Figure \ref{fg:qso_tracks} also shows the IRAC color tracks as a
function of redshift for three classes of object: pure AGNs, objects
with 50\% AGN and 50\% galaxy, where the percentages represent the
contribution to the total ``bolometric luminosity'' (as defined in
\citetalias{assef09}), and pure galaxies. Pure AGNs fall inside the
region defined by \citet{stern05} at most redshifts regardless of
reddening, with the exception of quasars with low reddening at
$z\simeq 4.5$ and $z\gtrsim 7$, where $\rm{H}\alpha$ and
$\rm{H}\beta$, respectively, are redshifted into the IRAC [3.6]
channel making the colors overly blue \citepalias[for a detailed
discussion about this point see \S4.3 of][]{assef09}. Objects with
combined AGN and galaxy components of equal luminosity
(Fig. \ref{fg:qso_tracks}c) have colors that generally stay inside,
but close to, the AGN selection boundary. Increasing the host galaxy
component will take these objects out of the selection region,
especially when considering the smaller region used for optically
extended AGNs. This is a bias against black holes accreting at low
Eddington ratios, with the accretion limit up to which these selection
criteria would target a given object being significantly dependent on
the reddening of the central source and the overall SED of the
source. For $z\lesssim 3$, the IRAC colors of galaxies are, by
construction, strictly outside the AGN selection boundaries. At higher
redshifts, galaxy colors overlap with those of active nuclei, but our
sample is too shallow to include non-active galaxies at these
distances.

\begin{figure}
  \begin{center}
    \plotone{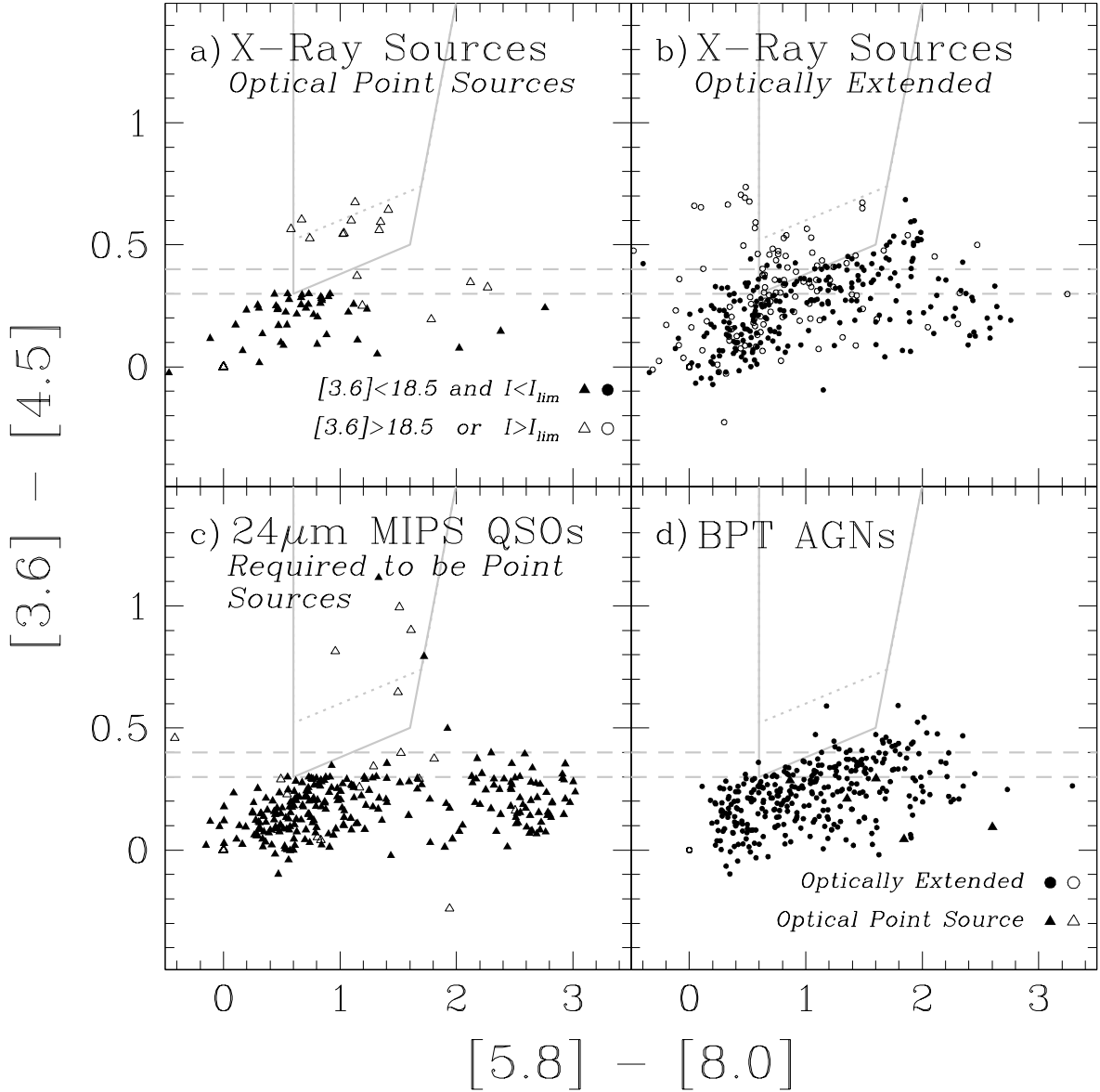}
    \caption{IRAC colors of spectroscopically observed objects with
    AGN signatures but {\it{not}} targeted as such based on their IRAC
    colors. We show objects targeted as X-ray sources, separated into
    optical point sources ({\it{top left}}) and optically extended
    objects ({\it{top right}}), as well as objects targeted as
    24$\mu$m MIPS QSOs ({\it{bottom left}}). We also show objects
    classified as AGNs by their spectral line ratios in the BPT
    diagram ({\it{bottom right}}). In all panels, solid symbols show
    sources bright enough to make it into the IRAC selected samples,
    namely $[3.6]<18.5$ and $I<I_{\rm lim}$, while open symbols show
    sources fainter than either of these limits. Note that $I_{\rm
    lim} = 21.5$ for optical point sources ({\it{triangles}}) and
    $I_{\rm lim} = 20$ for the optically extended objects
    ({\it{circles}}).}
    \label{fg:non_irac_agn}    
  \end{center}
\end{figure}

The bias against quasars that are faint in relation to their hosts is
further confirmed by Figure \ref{fg:non_irac_agn} \citep[see
also][]{hopkins09}. This Figure shows the IRAC colors of objects that
do not meet the IRAC color criteria but have AGN signatures based on
some other selection method. These correspond to objects that were
either targeted by AGES as X-ray or 24$\mu$m sources \citepalias[see
\S2 of][]{assef09}, or that show narrow line ratios typical of type II
AGN \citep[at higher ratios than the maximum star formation line
of][]{kewley01} in the BPT diagram \citet{baldwin81} classifications
of \citet{moustakas09}. We limit the X-ray sources to objects with at
least 4 counts in XBo\"otes because incompleteness is a significant
problem below this limit \citep{kenter05}. The IRAC criteria selects
almost all optically unresolved X-ray sources brighter than the
$[3.6]\leq 18.5$ magnitude limit, but misses large numbers of $24\mu$m
MIPS sources and optically extended X-ray sources. The great majority
of the MIPS sources correspond to star-forming galaxies at $z\sim 0.7$
that are compact enough to appear as optical point sources, but 46 of
them show clear signs of AGN activity (see
\S\ref{sec:mips_selection}). Non-IRAC selected optically extended
X-ray sources are likely real AGNs whose hosts dominate their IRAC
colors \citep[see][]{gorjian08}. This is primarily a problem at low
redshifts, as high redshift sources are too bright to have SEDs
dominated by their host galaxy.

As mentioned previously, unreddened AGNs at $z\sim 4.5$ have colors
too blue to enter the sample. In particular, in the redshift range
$4.10\lesssim z \lesssim 4.95$, the IRAC-selected AGNs sample is
highly incomplete, as it is too shallow to include reddened AGNs at
these redshifts that would meet the selection criteria. The problem is
similar to that of optical surveys at $2.5\lesssim z \lesssim 3.5$
\citep{fan99}, although here we are avoiding contamination by galaxies
rather than stars. In fact, the MIPS selection criteria identified
several quasars in this redshift range that were missed by the IRAC
selection criteria because of their blue colors. In order to address
the biases at low and high redshifts we have described, we include in
our catalog all X-ray and MIPS selected AGNs with magnitudes
$[3.6]<18.5$ and $I<21.5$ that are not flagged as such by their IRAC
colors. In the following sections we describe these selection criteria
in detail and how they complement our main IRAC-selected AGN sample.

\subsection{MIPS Selection}\label{sec:mips_selection}

The AGES MIPS quasar candidates are defined as optical point sources
with 24~$\mu$m MIPS fluxes $F_{24\mu \rm m} > 0.3$~mJy and that have
3\arcsec\ I-band aperture magnitudes $I > 18 - 2.5~\log(F_{24\mu \rm
m}/\rm {mJy})$. The limit on the I-band magnitude eliminates normal
stars. The largest contaminants to this group are low redshift,
strongly star-forming but optically unresolved galaxies, and
intermediate redshift ($z\sim 0.7$) star-forming galaxies that are
too distant to be resolved in the NDWFS survey. We also require that
$I<21.5$ and $[3.6]<18.5$ in order to match the IRAC magnitude limits.

There are 264 objects targeted as AGN by their MIPS fluxes and optical
morphology but not by their IRAC colors. As was done with the
IRAC-selected objects in the previous section, we eliminate all
objects that do not show broad emission lines in their spectra and
that are not classified as Type II AGN by \citet{moustakas09}. Under
this criteria, we add 46 sources to our sample, of which 35 are at
$z\leq 1$, 11 are at $z>1$ and seven are at $z>4$. We note that
although the AGN selection efficiency is very low for this sub-sample
(only 17\% are real AGN), this is not the case for the full sample of
MIPS targeted AGNs in the AGES survey, as there is a large overlap
with the IRAC selection criteria. For example, if we take all objects
in our sample targeted as AGN by the MIPS criteria, including the
IRAC-selected ones, we find 1401 objects of which 968 (69\%) are real
AGNs.

\subsection{X-Ray Selection}\label{sec:xray_selection}

The X-ray AGN candidates in the AGES survey correspond to all optical
sources matched with a probability of at least 25\% to an X-ray source
of 2 or more counts in the XBo\"otes survey according to the matching
approach of \citet{brand06}. The main contaminants are X-ray active
stars along with a small number of very low redshift galaxies. To
avoid stars, we eliminate all objects with redshifts below
$z=0.001$. We eliminate all objects detected with fewer than 4 counts,
because below this flux the incompleteness of the XBo\"otes survey is
high \citep{kenter05}. As done with the MIPS selection criteria, we
require that $I<21.5$ and $[3.6]<18.5$ in order to match the IRAC
magnitude limits. We assume all objects selected in this way
correspond to real active nuclei. This adds 333 objects to our sample,
of which 304 are optically extended and 29 are optical point
sources. Of these objects, 330 have $z<1$ and only 3, all optical
point sources, have $z>1$.

\subsection{The Full AGN Sample}\label{ssec:full_sample}

\begin{deluxetable}{c c  c  r r r  c  r r r  c  c r c  c  r r r}

\tablecaption{AGN Sample\label{tab:num_sample}}
\tablehead{ & & & \multicolumn{3}{c}{All (1838)} & &
             \multicolumn{3}{c}{IRAC (1459)} & & \multicolumn{3}{c}{MIPS\tablenotemark{a} (46)} & &
             \multicolumn{3}{c}{X-RAY\tablenotemark{b} (333)} \\
	     & & & \multicolumn{1}{c}{Ext.} & \multicolumn{1}{c}{P.S.} & \multicolumn{1}{c}{Total} &
             & \multicolumn{1}{c}{Ext.} & \multicolumn{1}{c}{P.S.} & \multicolumn{1}{c}{Total} & & 
	     \multicolumn{3}{c}{P.S.}  & & \multicolumn{1}{c}{Ext.} & \multicolumn{1}{c}{P.S.} & 
	     \multicolumn{1}{c}{Total}}
\tabletypesize{\small}
\tablewidth{0pt}
\tablecolumns{14}

\startdata 

All $z$ & $I<20\phd\phn$ & & 344 &  585 &   929 & &  40 &  557 &  597  & & \phn\phn &  18  & & & 304 &   10 &  314 \\
        & $I<21.5$       & & 344 & 1494 &  1838 & &  40 & 1419 & 1459  & & &  46  & & & 304 &   29 &  333 \\
\\
$z<1$   & $I<20\phd\phn$ & & 344 &  225 &   569 & &  40 &  203 &  243  & & &  13  & & & 304 &    9 &  313 \\
        & $I<21.5$       & & 344 &  408 &   752 & &  40 &  347 &  387  & & &  35  & & & 304 &   26 &  330 \\
\\
$z>1$   & $I<20\phd\phn$ & &   0 &  360 &   360 & &   0 &  354 &  354  & & &   5  & & &   0 &    1 &    1 \\
        & $I<21.5$       & &   0 & 1086 &  1086 & &   0 & 1072 & 1072  & & &  11  & & &   0 &    3 &    3 \\

\enddata

\tablecomments{The table shows the number of objects in our sample
divided by selection criteria, redshift and optical extension
(Ext. for optically extended sources and P.S. for point sources). The
numbers in parenthesis correspond the total number of objects selected
by each criteria. See \S\ref{sec:QSO_selection} for details.}

\tablenotetext{a}{Only objects not selected as IRAC AGNs.}
\tablenotetext{b}{Only objects not selected as either IRAC or MIPS AGNs.}

\end{deluxetable}

\begin{figure}
  \begin{center}
    \epsscale{0.85}
    \plotone{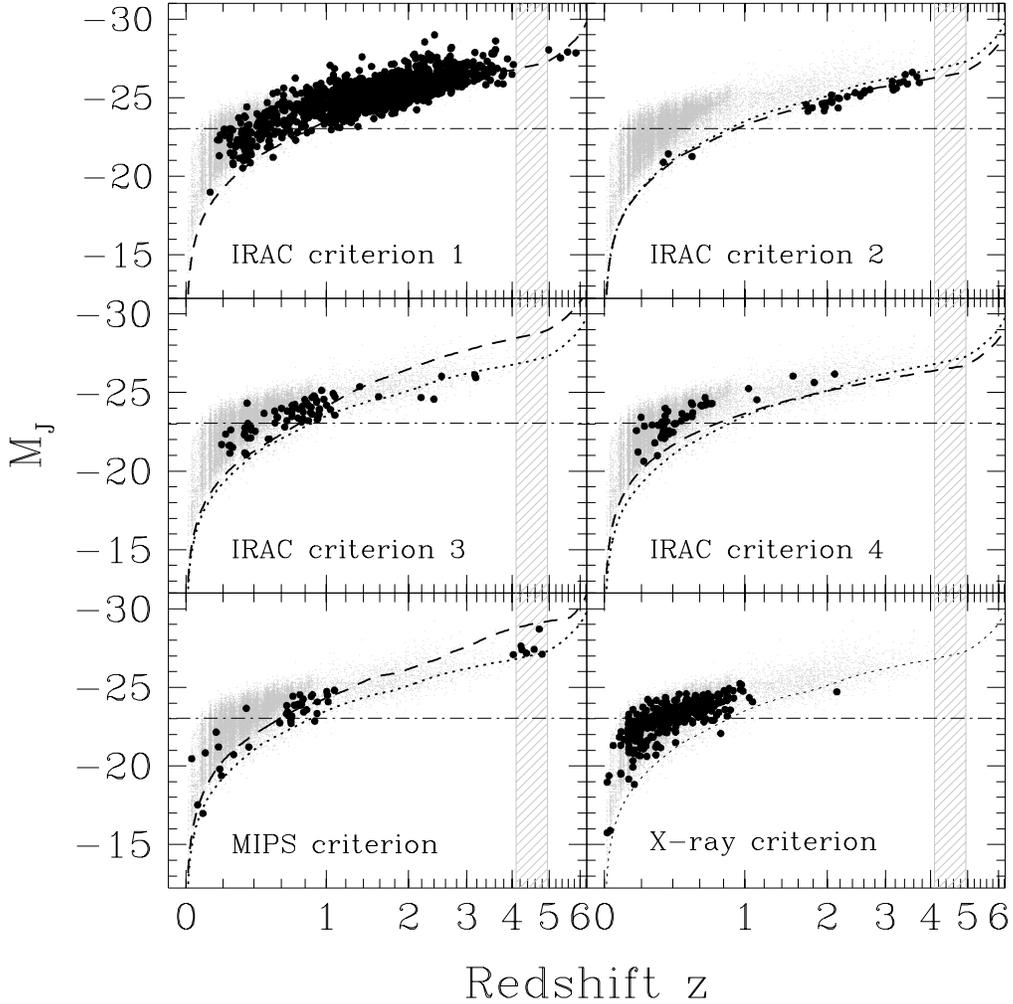}
    \caption{Magnitude-redshift relation of the sample used to derive
    the QSO luminosity function. Each panel shows one of the AGN
    targeting criteria of the AGES survey discussed in the text. Solid
    black circles show the AGNs targeted by each criteria while the
    gray dots show all other objects (except stars) in the survey,
    primarily normal galaxies. The dashed line in each panel shows the
    approximate $J-$band magnitude limit for the corresponding
    selection criterion, as extrapolated from the I and [3.6] band
    magnitude limits and the mean SED of each selection method. For
    the MIPS criterion we also include the 24$\mu$m magnitude limit of
    $0.3~\rm mJy$. We do not estimate this limit for the X-ray
    selected objects as the strongest constraints are in the X-rays,
    outside the range of our templates. For reference, in each panel
    the dotted line shows the $J-$band magnitude limit of the IRAC
    criterion 1. AGNs below the extrapolated magnitude limits are
    caused by a combination of photometric errors and host
    contamination below that of the average object of each class (see
    text for details).  The vertical hashed regions shows the
    approximate redshift range on which our IRAC selected AGN sample
    contains no objects due to the contribution of the broad H$\alpha$
    emission line to the [3.6] channel. The dot-dashed black
    horizontal line shows the characteristic Schechter function
    $J-$band luminosity $M_{*,J}$ for local galaxies from
    \citet{cole01}.}
    \label{fg:n_agn_z}
    \epsscale{1}
  \end{center}
\end{figure}

Our final sample is described in detail in Table \ref{tab:num_sample}
and consists of 1494 optical point source and 344 optically extended
AGNs. Note that in this Table, and in the remainder of this paper, the
MIPS-selected sample refers to the sample of sources selected by their
MIPS fluxes and optical extension (see \S\ref{sec:mips_selection}) but
not by their IRAC colors. Similarly, the X-ray selected sample refers
to X-ray selected AGN not selected by either the MIPS or the IRAC
criteria. Figure \ref{fg:n_agn_z} shows the magnitude-redshift
relations of these sources. Notice that there are no IRAC-selected
sources at $4\lesssim z \lesssim 5$, but that this redshift range is
well populated by MIPS-selected QSOs, confirming the biases predicted
by our SED templates. Note too that five objects targeted by the IRAC
criterion 4 lie at $z>1$. These objects were incorrectly classified as
extended sources, and hence targeted by criteria 4 and 5 only. Upon
visual inspection of the optical images, four objects seem to have
slightly extended PSFs, although the nature of this is unclear, and
the other has a very close companion, which likely affected the
SExtractor stellarity index. We will consider them as optical point
sources for the rest of the analysis, although they are such a small
fraction of the objects that their treatment has little quantitative
impact. Figure \ref{fg:n_agn_z} also shows the approximate $J-$band
magnitude limits for each selection criterion using the mean SED
(normalized by bolometric luminosity) for each selection
criterion. Objects seen below these model magnitude limits are caused
by a combination of photometric errors and varying levels of host
contamination compared to that of the mean SED. In particular, high
redshift MIPS-selected objects are significantly fainter than the
magnitude limit because the mean SED is dominated by the more
numerous, host-dominated low-redshift sources. At low redshift,
MIPS-selected objects not targeted by an IRAC criterion have hosts
that are bright in relation to their AGN and hence the galaxy
dominates their SED. At high redshift, however, they are bright
quasars with very little host contamination to their SEDs that are
missed by the IRAC criteria because of the H$\alpha$ broad emission
line contribution to the [3.6] IRAC channel (see
\S\ref{ssec:IRAC_QSO_selection} for details). We account for all of
these problems in the selection functions by using the individually
observed SEDS to estimate them. Overall, our sample is still subject
to one type of bias, related to the maximum reddening of AGNs of a
given luminosity and host composition in our sample is a function of
redshift, due primarily to the optical flux limit of AGES
spectroscopy. The combination of IRAC, X-ray and MIPS targeting
criteria is, by itself, not strongly affected by this bias, up to the
point where the dust torus is self-obscuring in the mid-IR.

Notice that although we used the BPT diagram classification of
\citet{moustakas09} to separate AGN from inactive galaxies for the
IRAC and MIPS selection criteria, we do not include purely
BPT-selected AGNs. The main reason is that this sample is limited to
$z\leq 0.4$, and its inclusion would induce a systematic offset at
this transition redshift in the QLF. Also note that the selection
function of purely BPT-selected objects is more complex than for the
other samples, and it cannot be modeled with the tools we use here.

Given that our main reason for excluding purely BPT-selected AGNs is
the limitation of $z\leq 0.4$, we need to understand if using the BPT
diagram as a discriminator for contaminants in the IRAC- and
MIPS-selected samples introduces a bias. To test for biases due to not
being able to use the BPT method beyond $z=0.4$, we took the IRAC and
MIPS-selected AGNs solely confirmed as real AGNs by their BPT
classification (i.e. no X-ray detection and no broad emission lines),
fit their SEDs using the templates of \citetalias{assef09}, and
modeled their selection as IRAC or MIPS AGN at higher
redshifts. First, for the IRAC criteria, we find that the color of the
mean SED of these objects is inside the selection regions only at
$z\lesssim 0.6$ and $z\gtrsim 1.1$. Since we do not see non-active
galaxies in our sample at $z>0.8$ \citepalias[see][]{assef09} due to
the I-band limit of AGES, we can securely say that any bias introduced
would only affect the redshift range $0.4 < z < 0.6$. In fact, 93 of
the 114 IRAC-selected objects dropped in \S\ref{sec:QSO_selection} but
located inside the original \citet{stern05} selection diagram are at
$z<0.6$. For the MIPS criterion we find similar results. Of a total of
29 MIPS-selected AGN confirmed as real AGN solely due to their BPT
classification, 15 (52\%) would have met the selection criteria at
$z>0.4$ and only 2 (7\%) would do so at $z>0.6$. While it would be
necessary to understand the luminosity function of these objects to
assess the complete significance of this bias, these numbers suggest
that the largest effect is in the $0.4 < z < 0.6$ range and of little
importance at $z>0.6$.

\begin{figure}
  \begin{center}
    \plotone{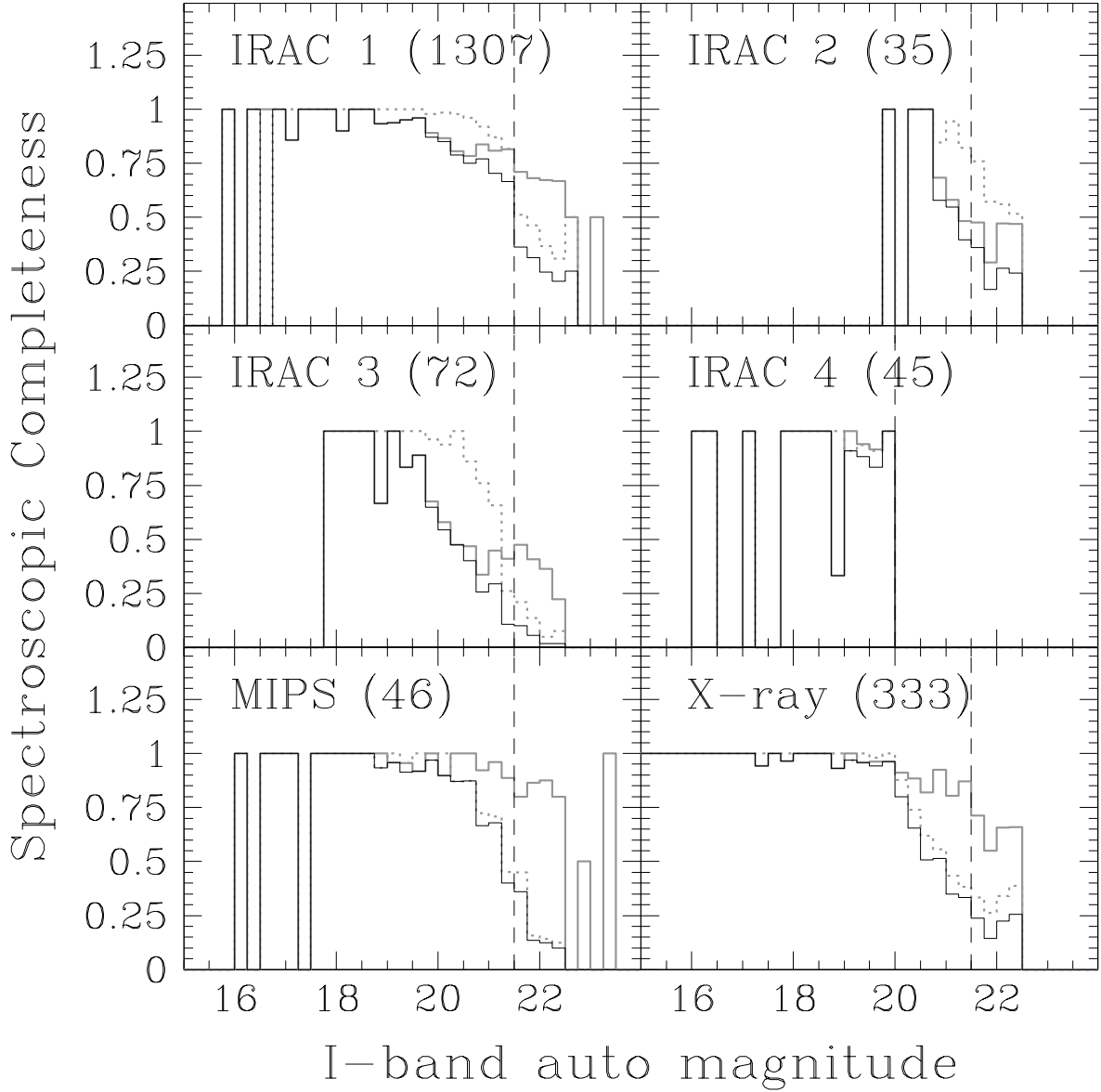}
    \caption{Spectroscopic completeness of our sample as a function of
    magnitude for each IRAC selection code discussed in
    \S\ref{ssec:qso_lf_det}. The solid gray line shows the fraction of
    objects as a function of I-band magnitude for which a spectrum was
    taken, while the dotted gray line shows the fraction of
    spectroscopically observed objects with a successful redshift
    measurement. The solid black line shows the product of the two,
    which is the inverse of the weight for an object of a given
    $I-$band magnitude when building the QLFs. The number in
    parenthesis is the number of objects in our sample from each
    category, and the vertical dashed line shows the magnitude limit
    to which we consider each selection criteria.}
    \label{fg:weights}
  \end{center}
\end{figure}

Our sample must be corrected for spectroscopic incompleteness in order
to estimate the QSO luminosity function, as not all objects targeted
for spectroscopy were observed, and redshifts could not be measured
for all the observed objects. Let $P_S(m_I)$ be the probability that a
spectrum of a given target was obtained, and $P_z(m_I)$ be the
probability that a redshift was successfully measured from that
spectrum, where both quantities are a function of the $I-$band
observed magnitude of the source, $m_I$. We correct our sample by
weighting each object by $[P_S(m_I) P_z(m_I)]^{-1}$. This weighting
assumes that the targets were randomly observed and that unsuccessful
redshift measurements are due solely to bad data. While the first
assumption is generally true due to the design and observing strategy
of the AGES survey, this is not completely clear for the second
one. In particular, redshift measurements become easier for objects
with stronger emission lines, which could in principle bias the
objects with unsuccessful redshift measurements to have lower AGN
luminosities. However, the QSO survey magnitude limit was $I=22.5$,
whereas we use a sample limited to a full magnitude brighter than this
limit, so it should not be a major source of bias. Figure
\ref{fg:weights} shows $P_S$ and $P_z$ as a function of observed
$I-$band magnitude, as well as the product of these two terms.

In the next section, we use this sample to study the QSO luminosity
function and its evolution with redshift. At $z<0.75$ a significant
number of objects are extended, and hence we use our full sample to
determine the QLF. However, IRAC-targeted extended sources (criterion
4 of \S\ref{ssec:IRAC_QSO_selection}) are strictly bounded by $I_{\rm
auto}\leq 20$, and they represent 10\% of the extended objects in the
sample up to this magnitude limit. In order to have a complete sample,
we limit our analysis at $z<0.75$ only to objects brighter than this
limit. For higher redshift bins, we use only optical point sources
limited to $I_{\rm auto} \leq 21.5$.

\section{The mid-IR and X-ray Selected AGN Luminosity Function}\label{sec:qso_lfs}

In this Section we will study the luminosity function of AGNs in our
sample. The combined data set described in \S\S\ref{sec:data} and
\ref{sec:QSO_selection} is, by construction, well-suited to study the
evolution of nuclear activity over a broad range of luminosities and
redshift. The impact of each selection criteria upon the absolute
magnitude limits as a function of redshift are illustrated in Figure
\ref{fg:n_agn_z}. In this Section we will also discuss how our
measurements compare to those determined by other groups.

\subsection{Luminosity Function Determination}\label{ssec:qso_lf_det}

We constructed rest-frame $J-$band luminosity functions. We chose
$J-$band because it is less affected by dust reddening than bluer
bands and it overlaps with at least one of the near- to mid-IR bands
available for our sample at every redshift. We use the
\citet{page2000} variant of the $V/V_{\rm max}$ method
\citep{schmidt68} which constructs luminosity functions binned in
absolute magnitude and redshift by weighting each object by the volume
in which it could have been observed and still be part of our
sample. We estimate the number of objects in a given magnitude and
redshift bin as
\begin{equation}\label{eq:v_vmax}
\Phi(M_J,z)\ =\ \sum_i\ \left\{\ W_i \int_{M_{J,\rm min}}^{M_{J,\rm
max}} \int_{z_{\rm min}}^{z_{\rm max}}\ f_i(M_J^{\prime},z^{\prime})\
\frac{dV}{dz^{\prime}}\ dz^{\prime}\ dM_J^{\prime}\right\}^{-1}
\end{equation}
\noindent where $M_{J,\rm min}$, $M_{J,\rm max}$, $z_{\rm min}$ and
$z_{\rm max}$ are the edges of the magnitude and redshift bin
respectively centered at $M_J$ and $z$, $V$ is the co-moving volume
and $W_i$ is the spectroscopic completeness of objects targeted by the
same IRAC criteria as object $i$ (as detailed in
\S\ref{sec:QSO_selection}). The function
$f_i(M_J^{\prime},z^{\prime})$ is the probability that object $i$
would have entered our sample if located at redshift $z^{\prime}$ with
a J-band absolute magnitude $M_J^{\prime}$. This function considers
all the selection criteria for each object, and not just the one by
which it was targeted. For each object, we estimate its best-fit SED
from the templates presented in \citetalias{assef09} and then
calculate the flux in each band assuming a redshift $z^{\prime}$ and a
J-band magnitude $M_J^{\prime}$ to determine if the object would have
passed the IRAC or MIPS selection criteria. If it did, then
$f_i(M_J^{\prime},z^{\prime}) = 1$. If not, but was detected by the
X-ray criteria, $f_i(M_J^{\prime},z^{\prime})$ takes the value of the
X-ray detection probability, discussed in detail in Appendix
\ref{sec:xray_det}. Otherwise, we set $f_i(M_J^{\prime},z^{\prime}) =
0$. Note that, when calculating $f_i(M_J^{\prime},z^{\prime})$, we
correct the estimated [3.6] and $I-$band magnitudes by the difference
between their 3\arcsec\ aperture and the auto magnitude
values. Because many of the sources are at high redshift, we have
added a prior based on the bright-end of the Las Campanas Redshift
Survey \citep{lcrslf} galaxy luminosity function to regularize the SED
fits and avoid unphysically bright galaxy contributions to the SEDs
\citepalias[this is similar to the approach for photometric redshifts
we used in][]{assef09}. In practice, this means we maximize
the probability
\begin{equation}
  P\ \propto\ e^{-\chi^2/2}\ e^{-L_R/L_{*,R}} ,
\end{equation}
\noindent where the $\chi^2$ term measures the difference between the
data and the model SEDs only, and $L_{*,R}$ is taken from the results
of \citet{lcrslf}. Note that our main conclusions do not change if we
eliminate this prior.

While the SED templates of \citetalias{assef09} do not explicitly
include evolution corrections, evolution, which is largely just a
shifting balance of star formation rates with redshift, is
automatically included as evolution in the typical weights assigned to
the templates with redshift. For example, passive evolution, to first
order, is correctly modeled by reducing the average contribution of
the early-type template relative to the later-type templates to the
observed SEDs. Other evolutionary effects, like changes in the mean
metallicity, are not considered in our SEDs templates, but their
effects are also more subtle. We note, however, that in the higher
redshift bins, our sample is composed primarily of Type 1 QSOs, and
hence most evolutionary effects on the hosts would have small impacts
upon the observed SEDs.

\begin{figure}
  \begin{center}
    \plotone{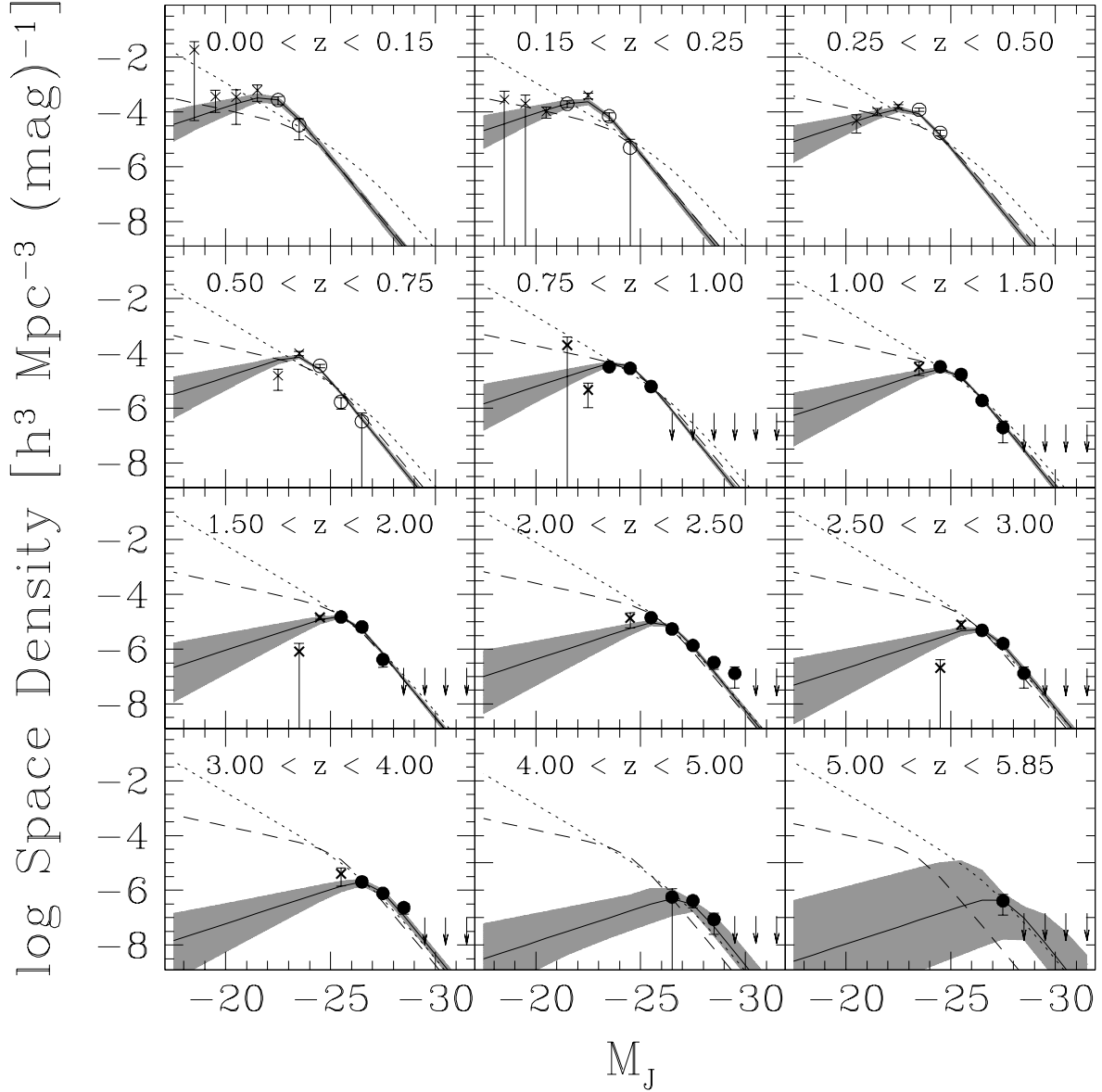}
    \caption{$J-$band luminosity function of our X-ray and mid-IR
    selected sample for several redshift bins. Here, the magnitudes
    have {\it{not}} been corrected for host contamination and
    reddening. At redshift $z<0.75$, the QLF is constructed from a
    combination of optically extended and point sources with $I<20$
    ({\it{open circles}}), while for $z>0.75$ it is constructed using
    only point sources with $I<21.5$ ({\it{solid circles}}). The
    crosses show points that were not used in the fits, as described
    in the text. The best-fit luminosity and density evolution (LDE),
    pure luminosity evolution (PLE) and pure density evolution (PDE)
    models are shown by the solid, dashed and dotted line
    respectively, although only the LDE model is an acceptable fit to
    the data. The shaded region shows the $2\sigma$ ($\Delta \chi^2
    \leq 4$) confidence region for the LDE fit.}
    \label{fg:lum_jband_lfs}
  \end{center}
\end{figure}

\begin{figure}
  \begin{center}
    \plotone{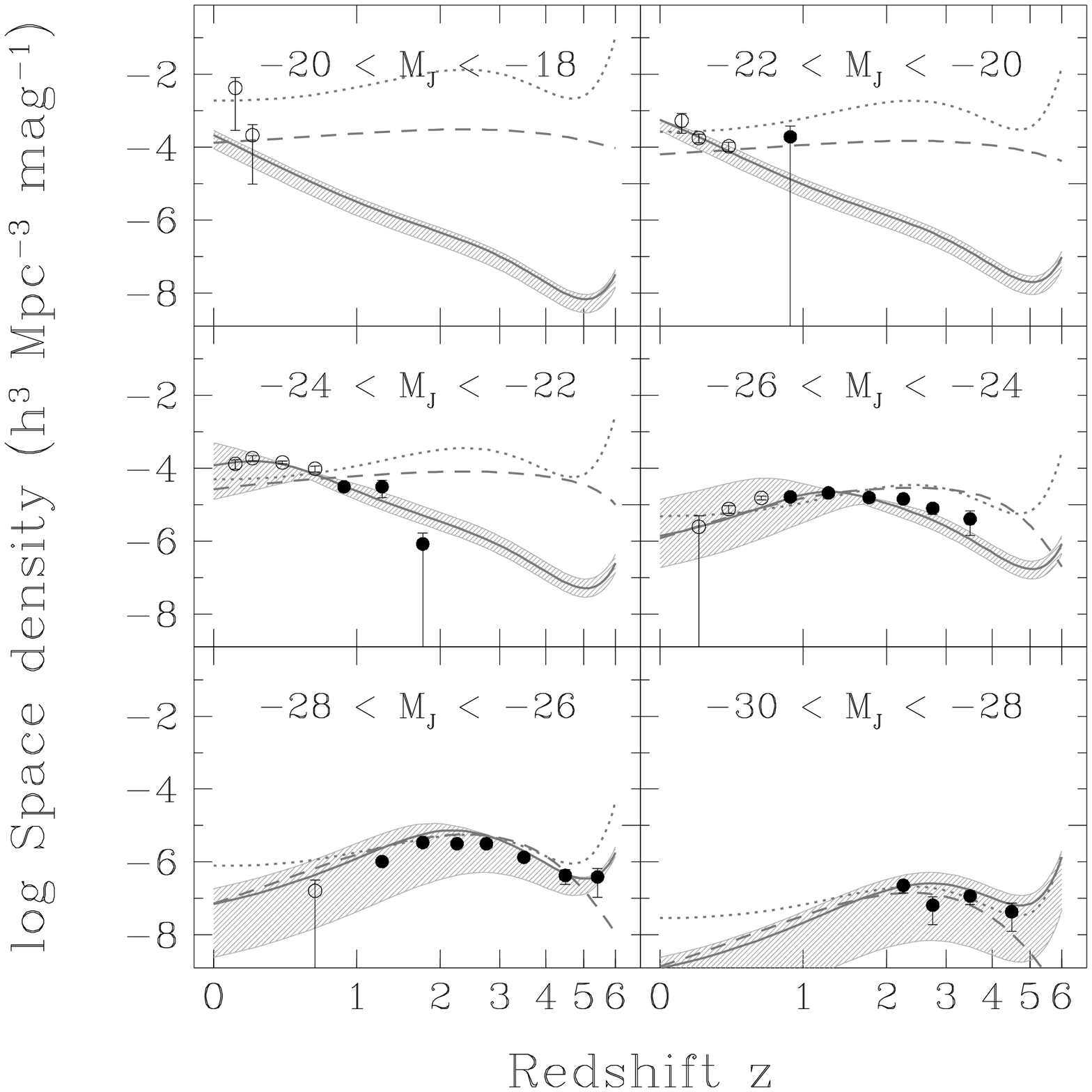}
    \caption{As in Figure \ref{fg:lum_jband_lfs}, but showing the
    space density of quasars as a function of redshift for several
    absolute magnitude intervals. The models are drawn based on the
    median magnitude of the objects in each $M_J$ bin. The hashed
    region shows the full range of space densities corresponding to
    each bin of our best-fit LDE model. The turn-over at $z>5$ is
    unlikely real.}
    \label{fg:z_jband_lfs}    
  \end{center}
\end{figure}

Figures \ref{fg:lum_jband_lfs} and \ref{fg:z_jband_lfs} show the QSO
luminosity functions for our complete sample. The data seems to be
consistent with a double power-law shape, where the bright end rises
towards faint magnitudes and the faint end is either constant or
falling towards fainter magnitudes. The uncertain behavior at the
faint-end is due to the small number of objects, and hence magnitude
bins, at all redshifts and the large dispersion between them. Doubling
the total number of magnitude bins at all redshifts does not modify
the observed behaviour. We warn the reader, however, that some of the
fainter magnitude bins may be unreliable due to the small amount of
objects and the large space density corrections due to the survey
limits. Also, the blended emission from the host galaxy affects the
overall shape of the luminosity function. Unlike traditional optical
QLFs, the host galaxy is a significant source of contamination in a
$J-$band QLF, as rest frame $J-$band lies close to the 1.6$\mu$m
stellar emission peak and the minimum of the AGN SED \citepalias[see
Figure 2 of][]{assef09}.

\begin{figure}
  \begin{center}
    \plotone{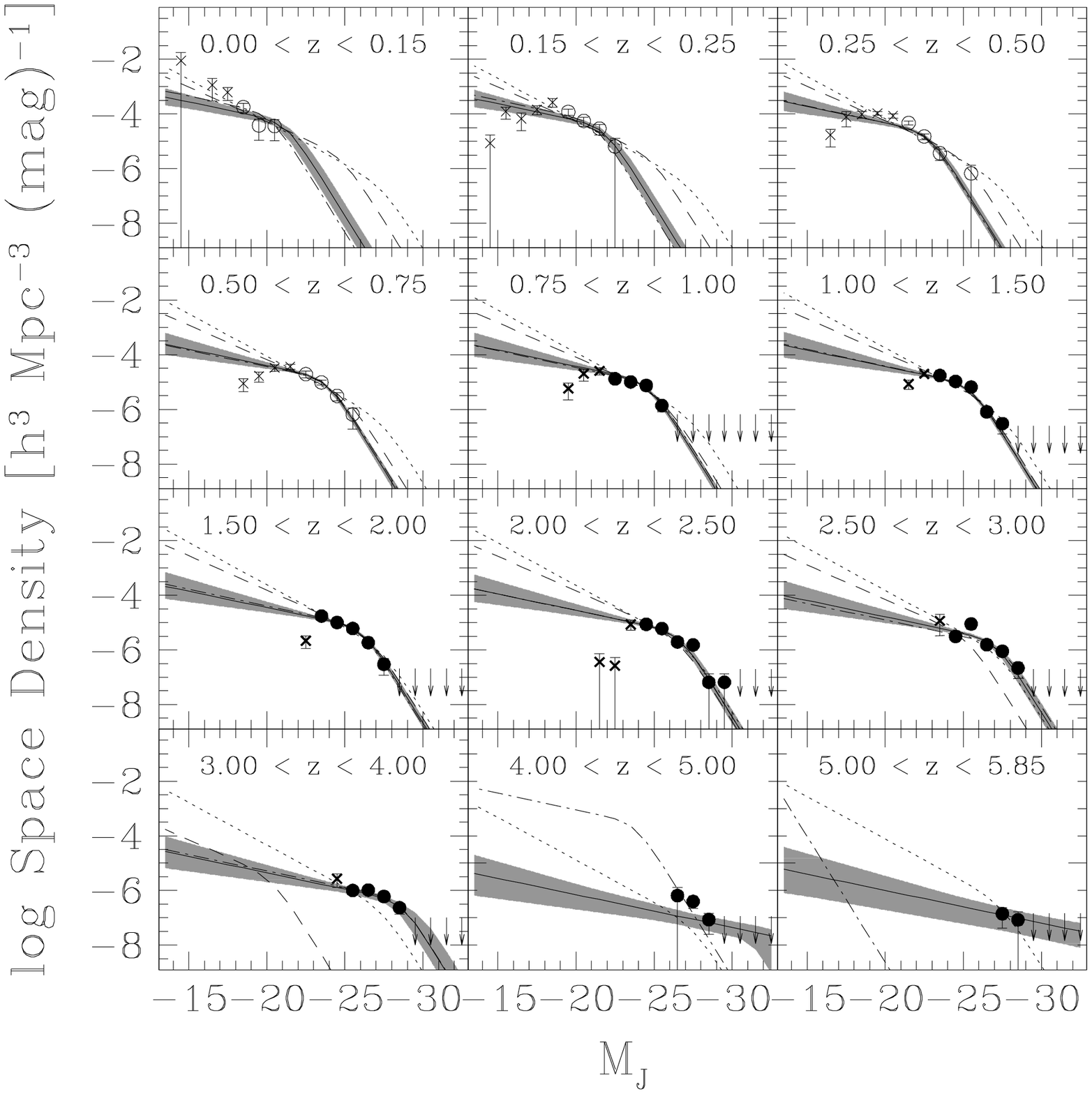}
    \caption{Same as in Figure \ref{fg:lum_jband_lfs} after removing
    host contamination. The dot-dashed line shows the best-fit
    allowing for an extra order in the $M_{*,J}(z)$ polynomial
    ($n=5$).}
    \label{fg:no_host_lum_lfs}    
  \end{center}
\end{figure}

\begin{figure}
  \begin{center}
    \plotone{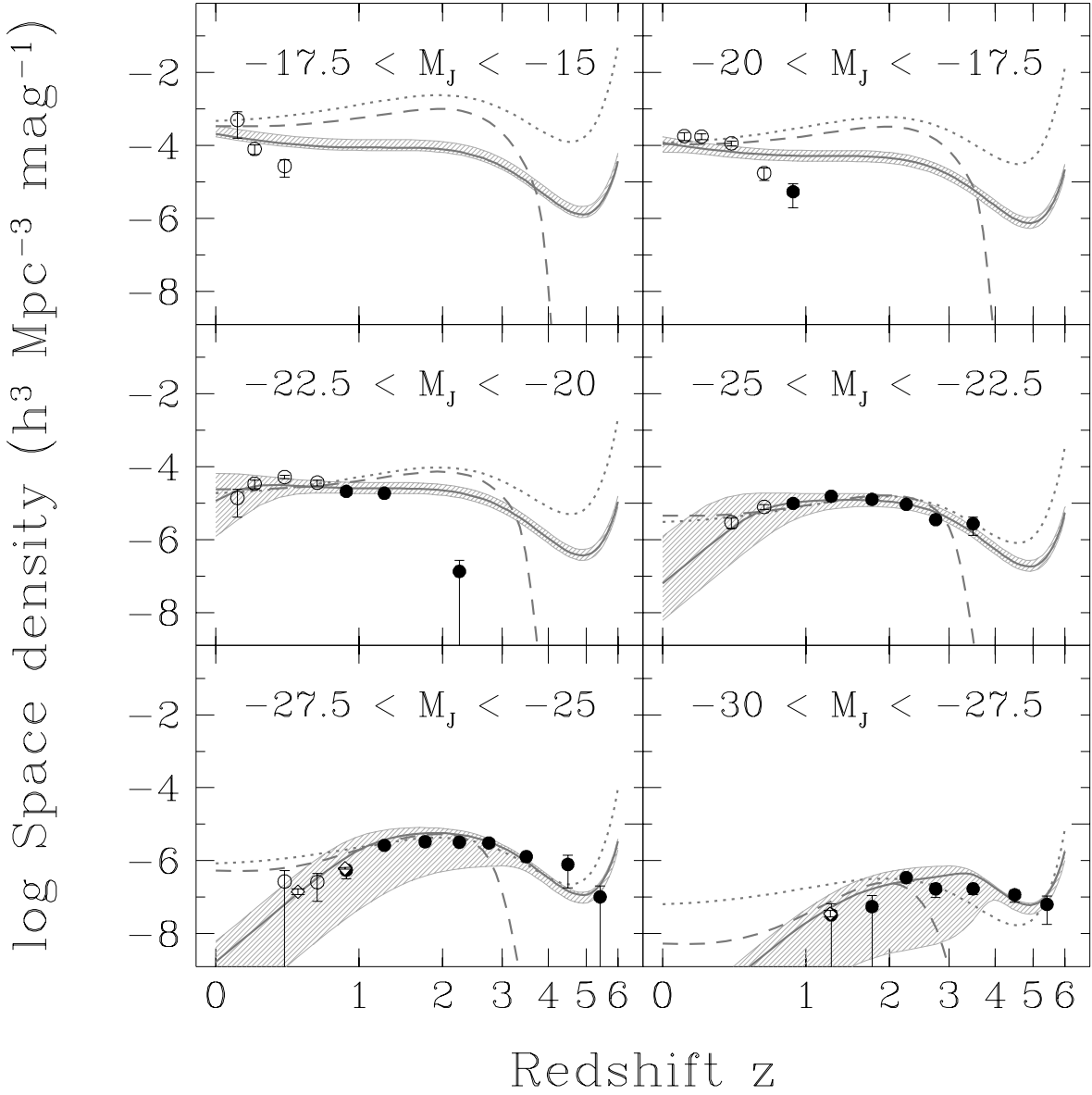}
    \caption{Same as in Figure \ref{fg:z_jband_lfs} after removing
    host contamination. The white rhombi in the two brightest
    magnitude bins show the lowest redshift space densities derived by
    \citet{richards06b} using the SDSS survey. The agreement is very
    good, as these points overlap with our measurements. Note that to
    convert $M_J$ to the redshift 2 $i-$band used by
    \citet{richards06b} we have used our AGN template with a reddening
    of $E(B-V)=0.05$, as discussed in \S\ref{sec:qso_lfs}. The upwards
    turn at $z>5$ is likely an artifact created by the small number of
    objects in the highest redshift bins coupled with the functional
    form of the model.}
    \label{fg:no_host_z_lfs}    
  \end{center}
\end{figure}

We use our SED templates to subtract the host contribution and
recalculate the QLF. We also correct the absolute magnitude for the
estimated reddening of the AGN, although this is not a major
correction at this wavelength --- the median reddening of
$E(B-V)=0.06$ changes the 1.2$\mu$m flux by only 5\%. Figures
\ref{fg:no_host_lum_lfs} and \ref{fg:no_host_z_lfs} show the resulting
host-corrected QLFs. Since this is a much more physically meaningful
representation of the QSO luminosity function, we will refer to it as
the ``true'' QLF for the rest of the paper. The luminosity functions
seems to be again best described by a double power-law. Due to the
small survey volume at low redshifts, the high luminosity power-law is
not present at $z<0.25$ because such objects are so rare. Unlike the
QLF found prior to removing the host contamination, the faint end
(without including the faintest magnitude bins), seems to be
increasing to fainter magnitudes, although with a much shallower slope
than that of bright quasars. Most redshift bins also show a
significant turnover at the faintest magnitudes. We do not believe
this to be real, but is again caused by small numbers of objects and
large space density corrections. Uncertainties in the host subtraction
may also play a role, although it is hard to quantify.

\begin{figure}
  \begin{center}
    \plotone{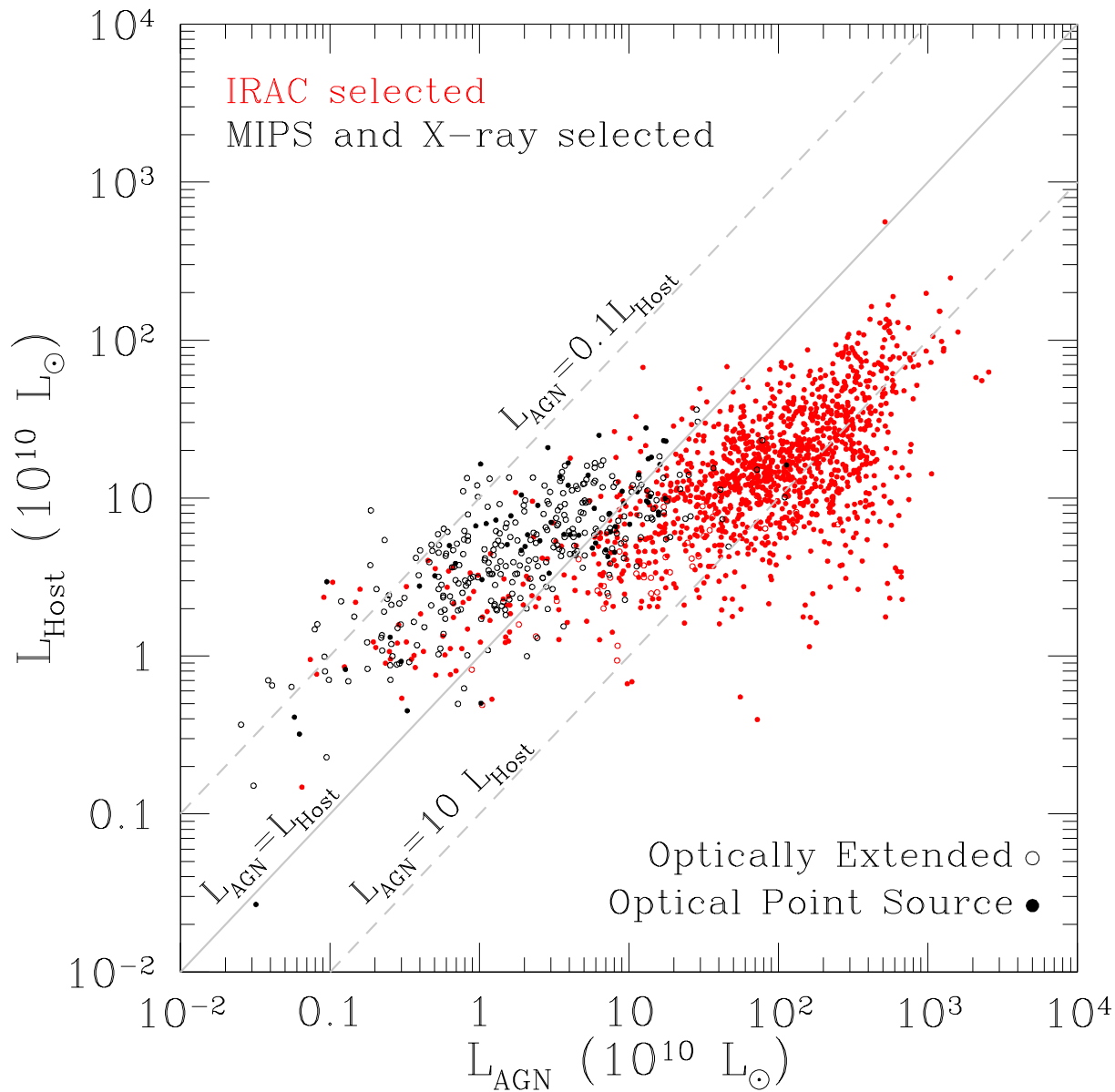}
    \caption{``Bolometric'' luminosity of the best-fit AGN component
    against that of the best-fit host component for all objects in
    our sample. Solid circles show optical point sources while open
    circles show optically extended objects. IRAC selected AGN are
    shown in red while MIPS and X-ray selected ones are shown in
    black.}
    \label{fg:Lagn_Lhost}
  \end{center}
\end{figure}

\begin{figure}
  \begin{center}
    \plotone{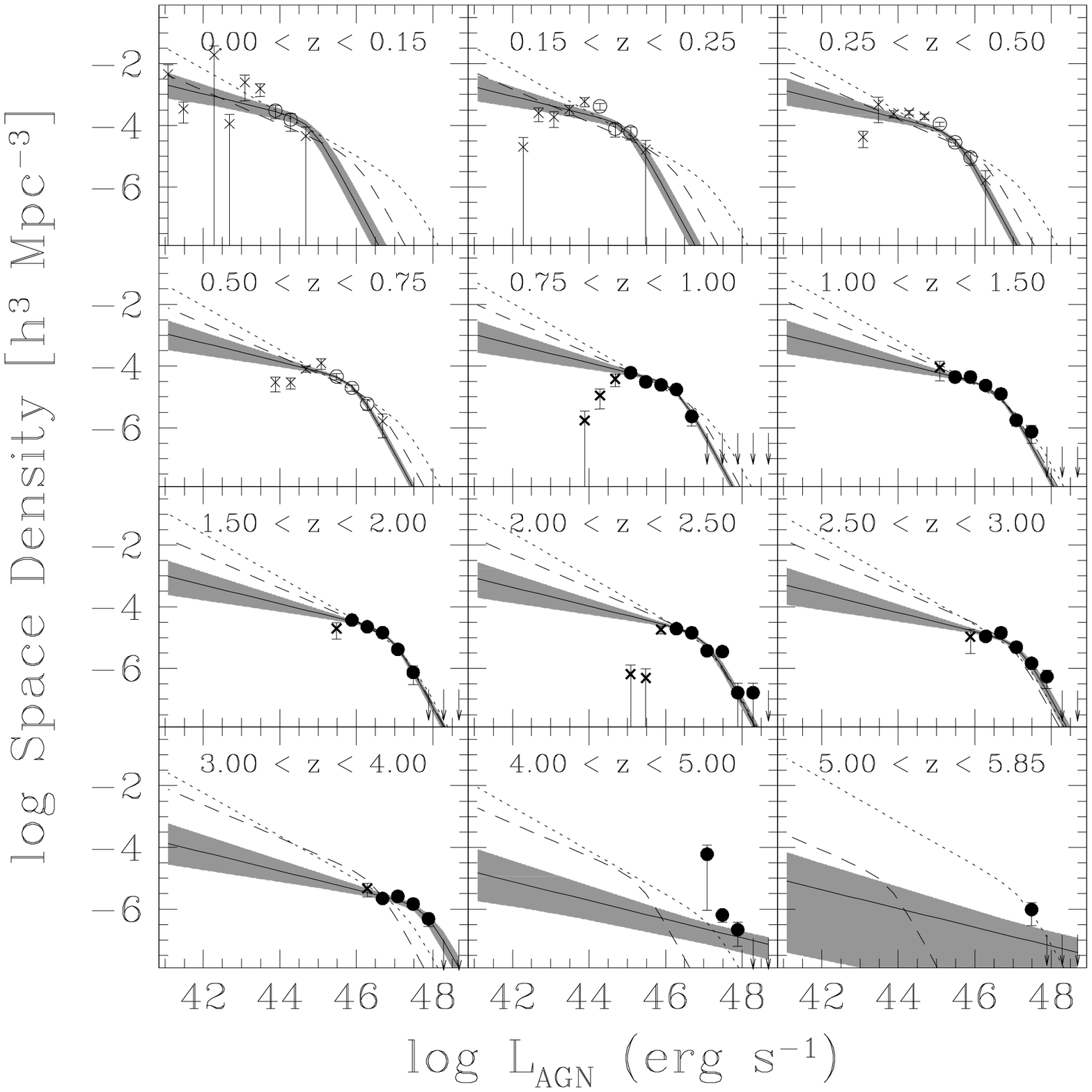}
    \caption{Bolometric QSO luminosity function. Lines and points have
    the same definition as in Figure \ref{fg:lum_jband_lfs}.}
    \label{fg:Lagn_lum_lfs}    
  \end{center}
\end{figure}

\begin{figure}
  \begin{center}
    \plotone{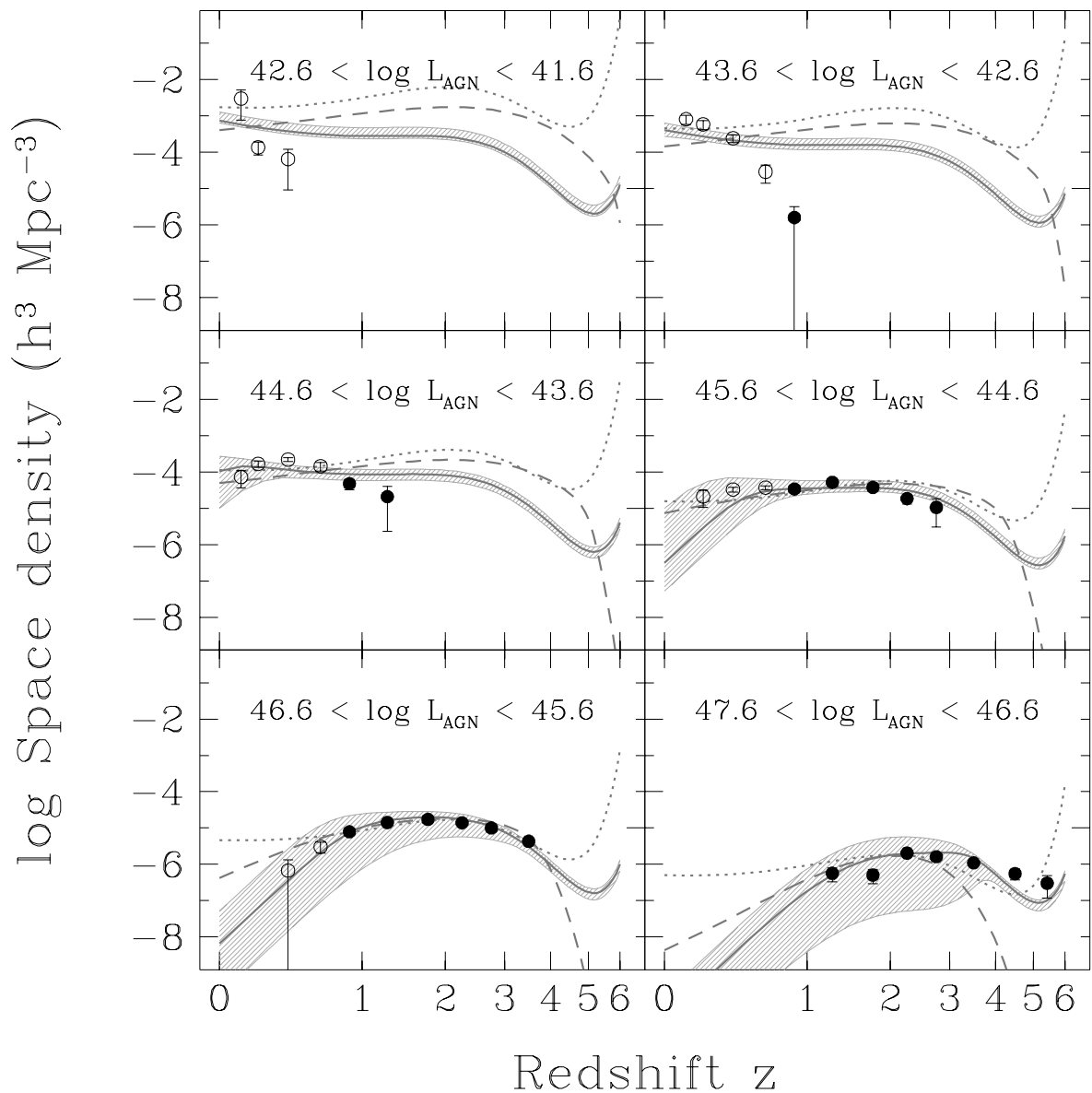}
    \caption{Redshift evolution of the bolometric QSO luminosity
    function for several magnitude bins. Lines and points have the
    same definition as in Figure \ref{fg:z_jband_lfs}. As for Figures
    \ref{fg:z_jband_lfs} and \ref{fg:no_host_z_lfs}, the upwards turn
    at $z>5$ is unlikely real.}
    \label{fg:Lagn_z_lfs}    
  \end{center}
\end{figure}

Just as we can use our template SED models to remove the host, we can
also estimate the ``bolometric'' luminosity function of QSOs. In this
case, the bolometric luminosity corresponds to the integrated light of
the unreddened best-fit AGN template to the observed SED of each
object over the wavelength range between 1216\AA\ and 30$\mu$m. The
QSO bolometric luminosity functions are shown for several redshift and
magnitude bins in Figures \ref{fg:Lagn_lum_lfs} and
\ref{fg:Lagn_z_lfs}. The bolometric and host corrected QLFs are very
similar since there is a simple relation between $L_{\rm AGN}$ and
$M_J$ for each individual source.

\subsection{Parametric fits to the QLF}\label{ssec:qlf_fit}

We model the QSO luminosity function (QLF) as a broken power-law,
\begin{equation}\label{eq:phi_mod}
\Phi(L_J,z)\ dL_J\ =\ \Phi_*(z)\
\left\{\left[\frac{L_J}{L_{*,J}(z)}\right]^{-\alpha} +
\left[\frac{L_J}{L_{*,J}(z)}\right]^{-\beta}\right\}^{-1}\
\frac{dL_J}{L_{*,J}(z)} ,
\end{equation}
\noindent where $\alpha$ and $\beta$ are free parameters and we allow
for both luminosity and density evolution. This functional form is
also used by, for example, \citet{croom04} and \citet{richards05} to
study the low redshift QSO luminosity function. While in principle the
shape of the QLF can be a function of redshift, we hold $\alpha$ and
$\beta$ fixed, as our sample lacks the high dynamic range in
luminosity at all redshifts that would be needed to explore this.

Note that our overall field size is small, so we have few high
luminosity quasars and our ability to constrain the bright-end slope
of the QLF is limited. This is not the case for $\beta$, as the depth
of the survey allows for a secure determination of this
parameter. This is in contrast to wide area shallow surveys, like 2QZ
\citep{croom04} and SDSS \citep{sdss}, that cover large areas to
shallow depths and can securely constrain $\alpha$ but not $\beta$. We
included a Gaussian prior on the bright-end slope $\alpha = -3.31\pm
0.05$ based on the results of the 2QZ survey. This estimate for
$\alpha$ comes from their fits using a redshift polynomial (see
eqn.[\ref{eq:mstar_others}] below) for the evolution of $M_{*}$ rather
than the model using the exponential evolution as a function of
look-back time (see \S\ref{ssec:qlf_comp}).

We model the break magnitude $M_{*,J}$ and its evolution by an $n^{\rm
th}$ degree Lagrange interpolation polynomial
\begin{equation}\label{eq:mstar_z}
M_{*,J}(z)\ =\ \sum_{i=1}^n M_{*,J}(z_i)\ \prod_{k=1, k\neq i}^n
\frac{(z-z_k)}{(z_i-z_k)},
\end{equation}
\noindent where the $M_{*,J}(z_i)$ are free parameters. This
parametrization is equivalent to the commonly used polynomial form
\citep[e.g.,][]{croom04,richards05,brown06}
\begin{equation}\label{eq:mstar_others}
M_*(z)\ =\ M_*(0) - 2.5 \sum_{j=1}^{n-1}\ k_j z^j , 
\end{equation}
\noindent but the $M_{*,J}(z_i)$ parameters are more physically
meaningful and have weaker covariances than the $k_j$ parameters, as
they simply represent the value of $M_{*,J}$ at different
redshifts. Similarly, we model the density evolution by
\begin{equation}\label{eq:phi_star_z}
\log[{\Phi_*(z)}]\ =\ \sum_{i=1}^n \log[{\Phi_*(z_i)}]\ \prod_{k=1,
k\neq i}^n \frac{(z-z_k)}{(z_i-z_k)}, 
\end{equation}
\noindent where all $\Phi_*(z_i)$ are free parameters. Note that,
because our sample covers a large spread in redshift around the peak
at $z\sim 2$, we must use $n\geq 4$ so that the evolution of the QLF
parameters can be asymmetric around the peak. In practice, we take
$n=4$ for modeling $M_{*,J}(z)$ and $n=5$ for modeling
$\log[{\Phi_*(z)}]$. For $M_{*,J}(z)$, $z_k = (0.5,1,2,4)$, while for
$\log[\phi_*(z)]$, $z_k = (0.25,0.5,1,2,4)$. The choice of the $z_k$
affects only the covariances of parameters --- the underlying
polynomial is independent of the choice and depends only on the
polynomial order.

We perform the fits for the three different cases of (1) luminosity
and density evolution (LDE, all 11 parameters are allowed to vary),
pure luminosity evolution (PLE, $\Phi_*(z)$ is replaced by a single
parameter $\Phi_*$) and pure density evolution (PDE, $M_{*,J}(z)$ is
replaced by a single parameter $M_{*,J}$). In the PLE and PDE models
there are 7 and 8 free parameters respectively. In all cases we use
the Levenberg -- Marquardt $\chi^2$ minimization algorithm including
the prior on the value of $\alpha$.

To fit the observed $J-$band QLF we drop faint end magnitude bins that
contain 2 or fewer objects. We also drop bins at all magnitudes that
have $V/V_{\rm max}$ corrections to their space densities greater than
a factor of 2, for $z<4$. Faint bins with so few objects and large
space density corrections are easily dominated by systematic errors in
the SED fits or in the $V/V_{\rm max}$ method. We note that for the
host corrected J-band luminosity function at $z<4$ there is only one
magnitude bin with 2 or less objects and a $V/V_{\rm max}$ correction
factor smaller than 2. This is the faintest bin of the $0.15 < z <
0.25$ redshift range, which is clearly well below any nearby data
point. At $z>4$, all magnitude bins are used in the fits, as there are
too few objects to do otherwise. We note, however, that all the $z>4$
bins have small correction factors ($<50\%$ corrections), except for
the faintest bin with $4<z<5$ of the host corrected $J-$band and
bolometric luminosity QLF. Removing this data point, however, changes
the best-fit parameters by much less than 1$\sigma$, and hence does
not play a fundamental role in determining the best-fit functional
form to the observed QLFs.

\begin{deluxetable}{l r r r}

\tablecaption{J-band QSO Luminosity Function Parametric Fits \label{tab:qsofit}}
\tablehead{Parameter & LDE & PLE & PDE}
\tabletypesize{\small}
\tablewidth{0pt}
\tablecolumns{4}

\startdata 

$\chi^2$                               &             49       &            155       &            180      \\
$\chi^2_{\nu}$                         &          2.053       &          5.536       &          6.657      \\
$\alpha$                               &  --3.35 $\pm$   0.05 &  --3.30 $\pm$   0.05 &  --3.30 $\pm$   0.05\\
$\beta$                                &  --0.37 $\pm$   0.27 &  --1.42 $\pm$   0.18 &  --2.13 $\pm$   0.08\\
$M_{*,J}     $                         &         \nodata      &         \nodata      & --26.71 $\pm$   0.37\\
$M_{*,J}(z = 0.5)$                     & --23.51 $\pm$   0.13 & --24.52 $\pm$   0.18 &         \nodata     \\
$M_{*,J}(z = 1.0)$                     & --24.64 $\pm$   0.10 & --25.16 $\pm$   0.18 &         \nodata     \\
$M_{*,J}(z = 2.0)$                     & --26.10 $\pm$   0.09 & --25.81 $\pm$   0.18 &         \nodata     \\
$M_{*,J}(z = 4.0)$                     & --27.08 $\pm$   0.22 & --25.10 $\pm$   0.21 &         \nodata     \\
$\rm{log}\ \phi_*$\tablenotemark{a}    &         \nodata      &  --4.53 $\pm$   0.11 &         \nodata     \\
$\rm{log}\ \phi_*(z = 0.25)$           &  --3.41 $\pm$   0.11 &         \nodata      &  --5.92 $\pm$   0.28\\
$\rm{log}\ \phi_*(z = 0.50)$           &  --3.73 $\pm$   0.08 &         \nodata      &  --5.83 $\pm$   0.26\\
$\rm{log}\ \phi_*(z = 1.00)$           &  --4.17 $\pm$   0.06 &         \nodata      &  --5.58 $\pm$   0.24\\
$\rm{log}\ \phi_*(z = 2.00)$           &  --4.65 $\pm$   0.05 &         \nodata      &  --5.14 $\pm$   0.24\\
$\rm{log}\ \phi_*(z = 4.00)$           &  --5.77 $\pm$   0.12 &         \nodata      &  --5.68 $\pm$   0.27\\
$M_{*,J}(0)$                           &      --22.05         &      --23.65         &         \nodata     \\
$k_1$                                  &         1.31         &         0.80         &         \nodata     \\
$k_2$                                  &       --0.30         &       --0.20         &         \nodata     \\
$k_3$                                  &         0.02         &         0.01         &         \nodata     \\

\enddata

\tablecomments{The best-fit parameters of our luminosity and density
evolution (LDE), pure luminosity evolution (PLE) and pure density
evolution (PDE) model fits to the QLF. For each parameter we quote the
formal 1$\sigma$ error-bars from the Levenberg -- Marquardt fitting
algorithm. For an easier comparison with other results in the
literature, we also show the values of the parameters of the more
commonly used functional form for the evolution of $M_*$ shown in
equation (\ref{eq:mstar_others}). We show no error-bars for this
values as correlations between them can be highly significant.}

\tablenotetext{a}{~$\rm h^3\ \rm Mpc^{-3}\ \rm mag^{-1}$}

\end{deluxetable}

\begin{deluxetable}{l r r r r r r}


\tablecaption{Host Corrected J-band QSO Luminosity Function Parametric Fits \label{tab:qsofit_nohost}}
\tablehead{Parameter & LDE & PLE & PDE & LDE--$\alpha(z)$ & LDE$_{\rm STY}$ & LDE$_{\rm STY}^\beta$}
\tabletypesize{\footnotesize}
\tablewidth{0pt}
\tablecolumns{7}

\startdata 

$\chi^2$                              &             62       &            154       &            137       &             61       & \nodata            & \nodata            \\
$\chi^2_{\nu}$                        &          1.630       &          3.677       &          3.341       &          1.605       & \nodata            & \nodata            \\
$\alpha_0$                            &         \nodata      &         \nodata      &         \nodata      &  --3.17 $\pm$   0.05 & \nodata            & \nodata            \\
$\alpha$                              &  --3.30 $\pm$   0.05 &  --3.31 $\pm$   0.05 &  --3.31 $\pm$   0.05 &         \nodata      &  --3.30            &  --3.30            \\
$\beta$                               &  --1.30 $\pm$   0.08 &  --1.63 $\pm$   0.06 &  --1.77 $\pm$   0.05 &  --1.26 $\pm$   0.09 &  --1.15 $\pm$ 0.05 &  --1.30            \\
$M_{*,J}     $                        &         \nodata      &         \nodata      & --27.51 $\pm$   0.34 &         \nodata      &  \nodata           &  \nodata           \\
$M_{*,J}(z = 0.5)$                    & --23.32 $\pm$   0.21 & --25.12 $\pm$   0.24 &         \nodata      & --23.15 $\pm$   0.25 & --22.68 $\pm$ 0.13 & --23.29 $\pm$ 0.09 \\
$M_{*,J}(z = 1.0)$                    & --25.00 $\pm$   0.17 & --25.81 $\pm$   0.25 &         \nodata      & --24.98 $\pm$   0.18 & --24.50 $\pm$ 0.10 & --25.04 $\pm$ 0.09 \\
$M_{*,J}(z = 2.0)$                    & --26.44 $\pm$   0.18 & --26.81 $\pm$   0.25 &         \nodata      & --26.26 $\pm$   0.19 & --26.49 $\pm$ 0.10 & --26.66 $\pm$ 0.07 \\
$M_{*,J}(z = 4.0)$                    & --30.86 $\pm$   1.15 & --14.23 $\pm$   2.03 &         \nodata      & --31.18 $\pm$   1.78 & --28.49 $\pm$ 0.14 & --30.14 $\pm$ 0.21 \\
${\rm{log}}\ \phi_*$\tablenotemark{a} &         \nodata      &  --5.48 $\pm$   0.12 &         \nodata      &         \nodata      &  \nodata           &  \nodata           \\
${\rm{log}}\ \phi_*(z = 0.25)$        &  --4.46 $\pm$   0.12 &         \nodata      &  --6.52 $\pm$   0.21 &  --4.34 $\pm$   0.13 &  --4.46            &  --4.46            \\
${\rm{log}}\ \phi_*(z = 0.50)$        &  --4.72 $\pm$   0.09 &         \nodata      &  --6.41 $\pm$   0.18 &  --4.65 $\pm$   0.10 &  --4.67 $\pm$ 0.02 &  --4.65 $\pm$ 0.02 \\
${\rm{log}}\ \phi_*(z = 1.00)$        &  --4.99 $\pm$   0.08 &         \nodata      &  --6.16 $\pm$   0.16 &  --4.97 $\pm$   0.08 &  --4.96 $\pm$ 0.03 &  --4.89 $\pm$ 0.04 \\
${\rm{log}}\ \phi_*(z = 2.00)$        &  --5.22 $\pm$   0.08 &         \nodata      &  --5.89 $\pm$   0.16 &  --5.16 $\pm$   0.09 &  --5.33 $\pm$ 0.04 &  --5.19 $\pm$ 0.04 \\
${\rm{log}}\ \phi_*(z = 4.00)$        &  --7.06 $\pm$   0.27 &         \nodata      &  --6.98 $\pm$   0.20 &  --7.02 $\pm$   0.36 &  --6.75 $\pm$ 0.07 &  --6.88 $\pm$ 0.08 \\
$M_{*,J}(0)$                          &      --20.57         &      --24.93         &         \nodata      &      --19.97         &  --20.09           &  --20.54           \\
$k_1$                                 &         2.72         &       --0.17         &         \nodata      &         3.20         &     2.42           &     2.67           \\
$k_2$                                 &       --1.12         &         0.77         &         \nodata      &       --1.42         &   --0.75           &   --1.02           \\
$k_3$                                 &         0.18         &       --0.25         &         \nodata      &         0.23         &     0.09           &     0.15           \\

\enddata

\tablecomments{The best fit parameters of each of our three models for the QLF with the formal 1$\sigma$ error-bars from the Levenberg -- Marquardt fitting algorithm. For an easier comparison with other results in literature, we also show the values of the parameters of the more commonly used functional form for the evolution of $M_*$ shown in equation (\ref{eq:mstar_others}). We show no error-bars for this values as correlations between them can be highly significant. This Table also shows the parameters obtained by varying the prescription of the fitting algorithms to explore possible systematic errors in our results, as discussed in \S\ref{ssec:qlf_fit}. Best fit parameters under the heading LDE--$\alpha(z)$ correspond to those obtained fitting the binned LF with the LDE model but assuming an $\alpha$ that evolves according to the results of \citet{hopkins07}. Those under the headings LDE$_{\rm STY}$ and LDE$_{\rm STY}^\beta$ were obtained with the STY method either fitting for $\beta$ or fixing it to --1.3, respectively (see \S\ref{ssec:qlf_fit} for details).}

\tablenotetext{a}{~$\rm h^3\ \rm Mpc^{-3}\ \rm mag^{-1}$}

\end{deluxetable}

\begin{deluxetable}{l r r r}

\tablecaption{Bolometric QSO Luminosity Function Parametric Fits \label{tab:qsofit_Lagn}}
\tablehead{Parameter & LDE & PLE & PDE}
\tabletypesize{\small}
\tablewidth{0pt}
\tablecolumns{4}

\startdata 

$\chi^2$                            &             64       &            144       &            149      \\
$\chi^2_{\nu}$                      &          1.818       &          3.681       &          3.922      \\
$\alpha$                            &  --3.31 $\pm$   0.05 &  --3.30 $\pm$   0.05 &  --3.30 $\pm$   0.05\\
$\beta$                             &  --1.30 $\pm$   0.07 &  --1.54 $\pm$   0.06 &  --1.70 $\pm$   0.04\\
$\rm{log}\ L_{*}$\tablenotemark{a}  &         \nodata      &         \nodata      &   46.80 $\pm$   0.11\\
$\rm{log}\ L_{*}(z = 0.5)$          &   45.24 $\pm$   0.08 &   45.93 $\pm$   0.10 &         \nodata     \\
$\rm{log}\ L_{*}(z = 1.0)$          &   45.97 $\pm$   0.07 &   46.30 $\pm$   0.08 &         \nodata     \\
$\rm{log}\ L_{*}(z = 2.0)$          &   46.51 $\pm$   0.07 &   46.62 $\pm$   0.09 &         \nodata     \\
$\rm{log}\ L_{*}(z = 4.0)$          &   48.25 $\pm$   0.36 &   45.54 $\pm$   0.16 &         \nodata     \\
$\rm{log}\ \phi_*$\tablenotemark{b} &         \nodata      &  --5.01 $\pm$   0.10 &         \nodata     \\
$\rm{log}\ \phi_*(z = 0.25)$        &  --4.02 $\pm$   0.12 &         \nodata      &  --5.84 $\pm$   0.17\\
$\rm{log}\ \phi_*(z = 0.50)$        &  --4.31 $\pm$   0.09 &         \nodata      &  --5.79 $\pm$   0.15\\
$\rm{log}\ \phi_*(z = 1.00)$        &  --4.61 $\pm$   0.08 &         \nodata      &  --5.58 $\pm$   0.13\\
$\rm{log}\ \phi_*(z = 2.00)$        &  --4.80 $\pm$   0.08 &         \nodata      &  --5.28 $\pm$   0.12\\
$\rm{log}\ \phi_*(z = 4.00)$        &  --6.67 $\pm$   0.22 &         \nodata      &  --6.23 $\pm$   0.16\\

\enddata

\tablecomments{The best fit parameters of each of our three models for
the bolometric QLF with the formal 1$\sigma$ error-bars from the Levenberg --
Marquardt fitting algorithm.}

\tablenotetext{a}{~erg s$^{-1}$}
\tablenotetext{b}{~$\rm h^3\ \rm Mpc^{-3}$}

\end{deluxetable}

The fits are summarized in Tables \ref{tab:qsofit},
\ref{tab:qsofit_nohost} and \ref{tab:qsofit_Lagn}, respectively, for
the three versions of the QLF we introduced in the previous section:
the J-band QLF uncorrected for host contamination
(Figs. \ref{fg:lum_jband_lfs} and \ref{fg:z_jband_lfs}), the J-band
QLF corrected for host contamination (Figs. \ref{fg:no_host_lum_lfs}
and \ref{fg:no_host_z_lfs}), and the AGN bolometric luminosity QLF
(Figs. \ref{fg:Lagn_lum_lfs} and \ref{fg:Lagn_z_lfs}). The fits are
shown in each of the respective Figures, along with the 2$\sigma$
($\Delta \chi^2 \leq 4$) confidence region of the LDE fit. The
relatively small confidence region at the highest redshift bins in
spite of the of the few data points is driven by a combination of the
small error bars of many of the data points in combination with the
stiffness of the model constrained by the lower redshift bins. Note
that we do not count the bright-end upper limits in accounting for the
number of degrees of freedom. In all cases the LDE fit gives a much
better description of the data. In particular, for the host corrected
$J-$band QLF, the $\chi^2_{\nu}$ is larger by a factor of 1.9 for the
PDE model and by a factor of 2.1 for the PLE model. This strongly
argues that the nature of the QSO LF evolution is a combination of
changes in characteristic magnitude of the AGNs and changes in the
total number of these objects as a function of cosmic time, both
effects with similar strengths. This is compatible with the recent
results of \citet{croom09}, who observed a similar behavior in a
combined sample of SDSS and 2SLAQ QSOs (see \S\ref{ssec:qlf_comp}).

As discussed before, our sample lacks the high dynamic range to study
the redshift evolution of the power-law slopes, and hence we have
assumed them to be redshift independent in the LDE, PLE and PDE models
we fit. There is evidence that the bright-end slope may evolve with
redshift \citep[see, e.g.,][]{richards06b,hopkins07,croom09}, and
hence a fixed value of $\alpha$ may bias the rest of the LF
parameters. While we cannot fit for an evolutionary form of $\alpha$,
we can assess the effects its evolution may have on the rest of the
parameters. For this we modify the LDE model by including a
redshift-dependent $\alpha(z)$ that evolves according to the best-fit
functional form of \citet{hopkins07}, given by
\begin{equation}
  \alpha(z)\ =\ \frac{2 \alpha_0}{\left[(1+z)/(1+z_{\rm
    ref})\right]^{k_{\alpha,1}} + \left[(1+z)/(1+z_{\rm
    ref})\right]^{k_{\alpha,2}}},
\end{equation}
\noindent where $z_{\rm ref}\equiv 2$. We keep $k_{\alpha,1}$ and
$k_{\alpha,2}$ fixed to their best-fit values in \citet[][Table
3]{hopkins07} and we let $\alpha_0$ vary, subject to a Gaussian prior
based on the value in \citet{hopkins07}. We refer to this model as
LDE--$\alpha(z)$. Its best fit parameters are listed in Table
\ref{tab:qsofit_nohost} and all are within 1-$\sigma$ of those
obtained assuming a redshift independent bright-end slope, with a very
slight improvement in $\chi^2_{\nu}$ of 0.025. This shows that the
assumption of a non-evolving bright-end slope is not a significant
source of bias in our results.

The best-fit LDE model, as shown in Figure \ref{fg:no_host_z_lfs},
implies that the number of faint AGNs ($M_J > -20$) is roughly
constant in the redshift range from 3 to 0. The behavior of the
data-points at $M_J > -20$ is consistent at the lowest redshift bins
but seems to deviate for $z>0.5$, suggesting a decrease in the number
of AGNs. We cannot tell if this behavior is real or not, as these
points correspond to the faintest magnitude bins at a given redshift
range which we have argued before are the most likely to be affected
by systematic errors. The brighter objects ($M_J < -20$) have a
roughly constant density down to a certain redshift after which their
space density decreases rapidly to lower redshifts. The redshift of
this inflection points seems to increase for brighter magnitudes,
showing a very clear trend of downsizing of AGN activity. The data are
generally consistent with this picture, however the constraints are
weaker. The drop in space density of bright quasars with diminishing
redshift is consistent with the data, although the large error bars
limit the significance. These large error bars are a direct
consequence of the small area of the Bo\"otes field. However, if we
simply superpose the space densities measured from the shallower but
1000 times larger area SDSS survey \citep{richards06b} on our
estimates for the two brightest magnitude bins, we find excellent
agreement, confirming the strong evolution of bright quasars. To
convert between the different bands, we used our AGN SED template with
$E(B-V)=0.05$ (see \S\ref{ssec:qlf_comp}). We do not use these SDSS
data in fitting the LDE model shown in Figure \ref{fg:no_host_z_lfs}.

\begin{figure}
  \begin{center}
    \plotone{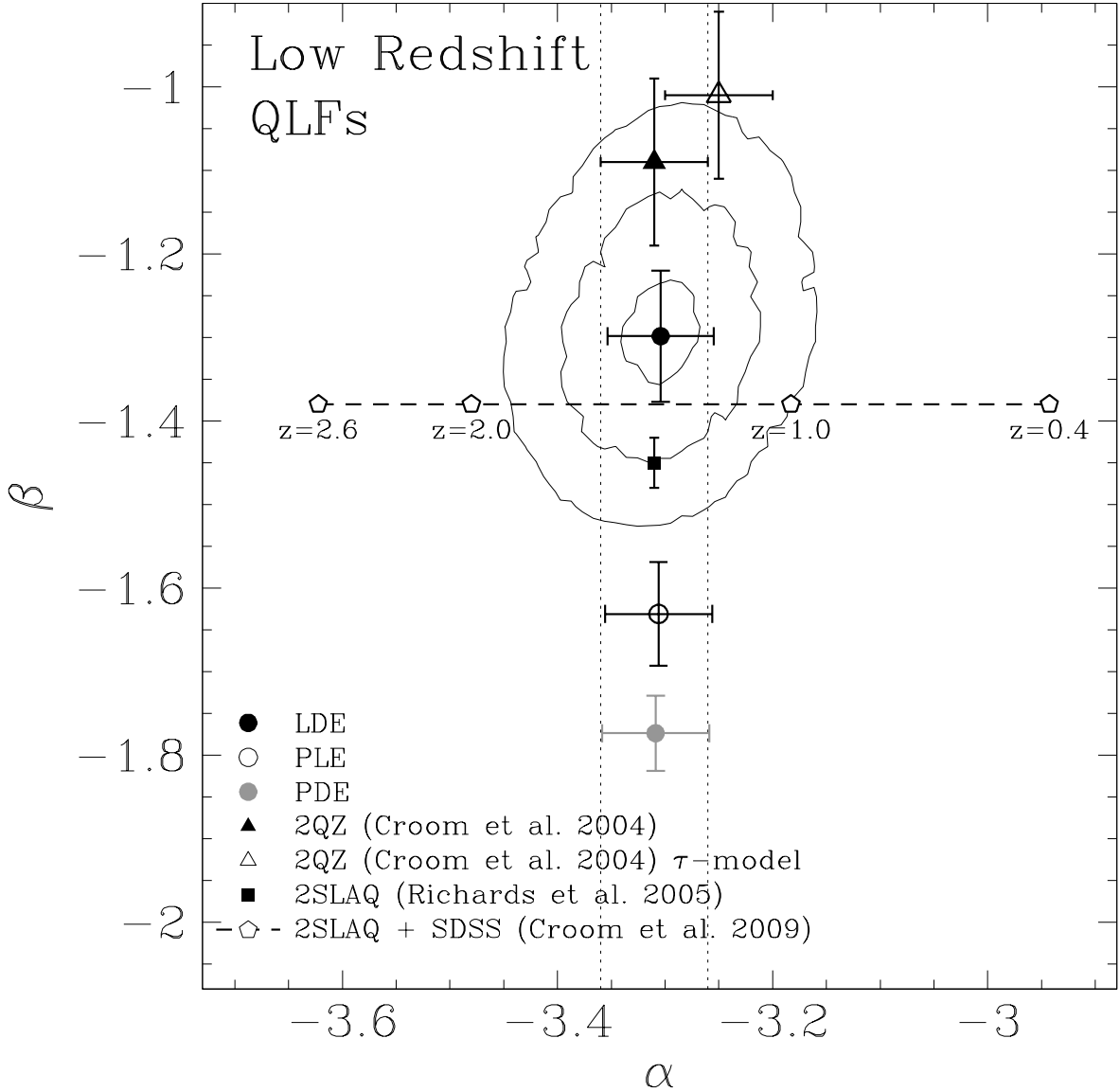}
    \caption{Best-fit values of the parameters $\alpha$ and $\beta$ to
    our data (see eqn. [\ref{eq:phi_mod}]) for the LDE ({\it{black
    solid circle}}), PLE ({\it{open circle}}) and PDE ({\it{gray solid
    circle}}) models. Error-bars are the formal errors determined by
    the Levenberg -- Marquardt minimization technique. Contours show
    regions of constant $\Delta \chi^2$ of 1, 4 and 9 ({\it{solid
    black line}}). The vertical dotted black lines show the 1$\sigma$
    interval of the prior used on the parameter $\alpha$. For
    comparison, we show values of other studies determined by using
    low redshift samples (see \S\ref{ssec:qlf_comp} for details). In
    particular, we show the best-fit values determined by
    \citet{croom04} using the 2QZ survey observations, by
    \citet{richards05} using early data from the 2SLAQ survey and by
    \citet{croom09} using a combination of the finalized 2SLAQ survey
    data and the QSO SDSS sample of \citet{richards06b}.}
    \label{fg:low_z_alpha_beta}
  \end{center}
\end{figure}

\begin{figure}
  \begin{center}
    \plotone{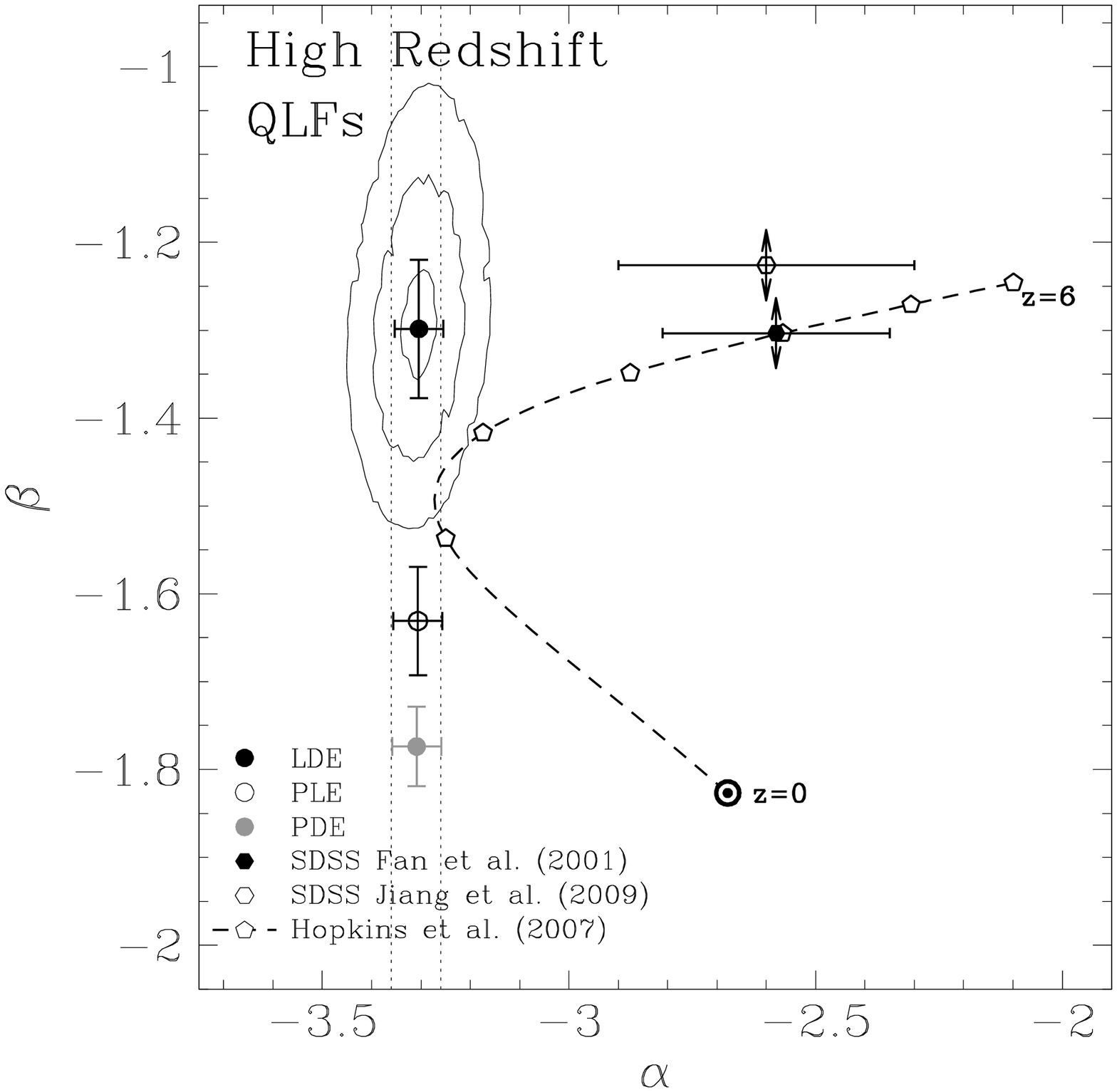}
    \caption{Same as Figure \ref{fg:low_z_alpha_beta}, but here we
    compare to parameters fitted to high redshift samples. In
    particular we show the best-fit $\alpha$ values of \citet{fan01}
    and \citet{jiang09} determined using a sample of high redshift
    QSOs from SDSS (see \S\ref{ssec:qlf_comp} for details). We also
    show the best-fit values of $\alpha$ and $\beta$ determined by
    \citet{hopkins07} as a function of redshift. The target shows the
    values at $z=0$ and then pentagons mark the different values at
    redshift intervals separated by $\Delta z = 1$ up to $z = 6$ (also
    see \S\ref{ssec:qlf_comp} for details).}
    \label{fg:high_z_alpha_beta}
  \end{center}
\end{figure}

Note that for the host-corrected QLF, the best-fits for all three
models have negative values of both $\alpha$ and $\beta$, unlike for
the uncorrected QLF. This is in much better agreement with what has
been found by several other studies. The fact that this change is
induced by the host contamination removal suggests that lower
luminosity AGNs tend to have lower $L_{\rm AGN}/L_{\rm Host}$ ratios,
or in other words, that they tend to be in relatively brighter host
galaxies. By removing the host component, the J-band absolute
magnitude of the faint objects become fainter, depopulating the region
around the magnitude break and populating the faint end of the QLF,
while the brightest part of the bright end is relatively
unaffected. Although there is no {\it{a priori}} reason to expect a
correlation between $L_{\rm AGN}$ and $L_{\rm AGN}/L_{\rm Host}$, we
see exactly this trend in Figure \ref{fg:Lagn_Lhost}, which compares
the ``bolometric'' luminosity of the host and AGN components of the
best-fit SED. Note that the detailed distribution of the sources is
subject to two biases. First, $L_{\rm AGN}/L_{\rm Host}$ must be large
enough for the source to be recognized as an AGN, but small enough for
the host luminosity to be measured from the SED fit, and second, we
can detect high luminosity sources over larger volumes. However,
neither of these should affect the general trend that the region of
$L_{\rm AGN}/L_{\rm Host}$ occupied by sources is a strong function of
$L_{\rm AGN}$. It is worth noting that \citet{hopkins05b} predicts,
based on simulations of galaxy mergers that include AGN feedback
processes, that the faint-end of the AGN luminosity function is
populated by massive black holes accreting at sub-Eddington rates
rather than by small black holes radiating at the Eddington limit,
much in line with what we find here.

\begin{figure}
  \begin{center}
    \plotone{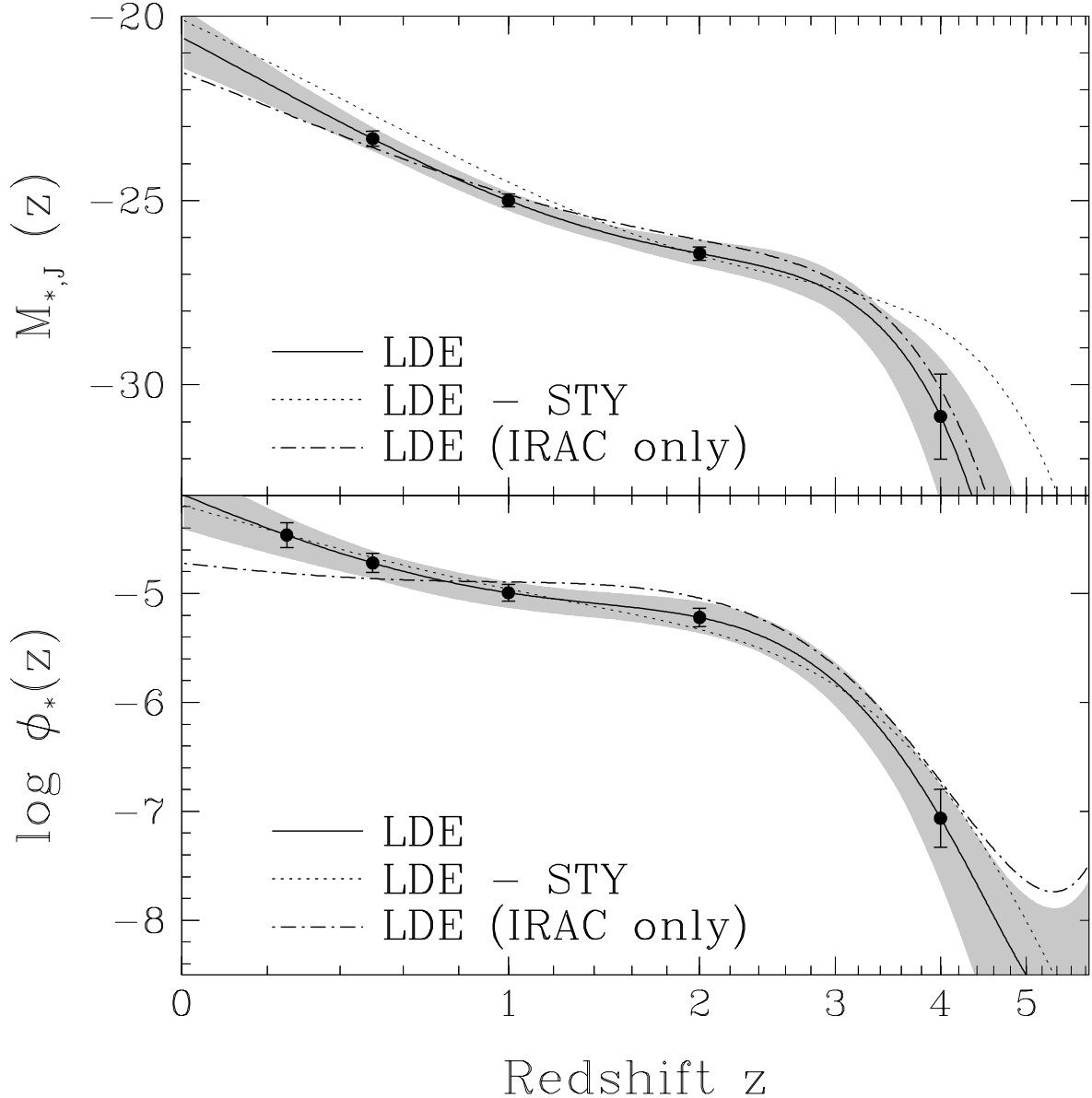}
    \caption{({\it{Top panel}}) The best-fit evolution of $M_{*,J}$
    from our LDE model for the full QSO sample and that obtained from
    our LDE model ({\it{solid line}}) for the IRAC selected sample
    only ({\it{dot-dashed line}}). We also show evolution obtained
    using the STY estimator to fit the LF ({\textit{dotted
    line}}). The gray shaded area shows the 2$-\sigma$ ($\Delta \chi^2
    \leq 4$) range spread of the evolution from the LDE fit. The
    points show the respective values of the best-fit $M_{*,J}(z_k)$
    parameters (see eqn. [\ref{eq:mstar_z}]) for this model and their
    respective error-bars from the Levenberg -- Marquardt fitting
    routine. ({\it{Bottom Panel}}) The best-fit evolution of
    $\phi_*$. The solid, dot-dashed and dotted lines, points and the
    gray shaded region have the same meaning as in the above panel.}
    \label{fg:evol}
  \end{center}
\end{figure}

Figures \ref{fg:low_z_alpha_beta} and \ref{fg:high_z_alpha_beta} show
our estimates for $\alpha$ and $\beta$, and their uncertainties, as
compared to other estimates (see \S\ref{ssec:qlf_comp} for
details). The error-bars in our measurements correspond to the formal
errors calculated from the Levenberg -- Marquardt fit, while the
contours show regions of constant $\Delta \chi^2$ (1, 4 and 9
respectively).  Figure \ref{fg:evol} shows the evolution of
$M_{*,J}(z)$ and $\phi_*(z)$ and their $2\sigma$ ($\Delta \chi^2 \leq
4$) uncertainties. We note that for $z<3$, the evolution of $M_{*,J}$
and $\log[{\Phi_*(z)}]$ is roughly linear with $\log(1+z)$, suggesting
that, at least in this range, $\log(1+z)$ would be a more natural
parameter to describe their evolution. At redshifts $z\geq 3$ the
evolution of both parameters show significant departures from
linearity in $\log(1+z)$, but this is not necessarily well
constrained. It is clear from Figure \ref{fg:no_host_lum_lfs} that the
highest redshift bins only weakly constrain these parameters, as all
the data points are fainter than $M_{*,J}$.

We experimented with changes in the order of the interpolation
polynomials from $n=4$ for the evolution of $M_{*,J}(z)$ and $n=5$ for
$\log[{\Phi_*(z)}]$. Reversing the number of parameters ($n=5$ for
$M_{*,J}(z)$ and $n=4$ for $\log[{\Phi_*(z)}]$) gives qualitatively
similar fits, albeit with a slightly worse $\chi^2_{\nu}$ for the LDE
model. Decreasing the number of parameters results in a worse fit to
the data as well. For example, $n=5$ for $\log[{\Phi_*(z)}]$ and $n=3$
for $M_{*,J}(z)$ results in an increase of $\chi^2_{\nu}$ of
approximately 0.2, while $n=4$ for both parameters produces and
increase of approximately 0.3. While $n=4$ for $M_{*,J}(z)$ and $n=5$
for $\log[{\Phi_*(z)}]$ do not fit the QLF with $\chi^2_{\nu}=1$, we
choose to not add more degrees freedom. Our LFs are somewhat noisy and
higher degrees of freedom will fit the noise. For example, in the host
corrected luminosity function, using $n=5$ for both parameters
improves the $\chi^2_{\nu}$ by $\approx 0.24$, better describing the
lower redshift bins at the cost of being unable to reproduce the
highest redshift measurements by a significant margin (see Figure
\ref{fg:no_host_lum_lfs}).

\citet{miyaji01} has argued that the method of \citet{page2000} may
bias the constructed luminosity function with respect to the true
underlying shape due to binning effects. To test for this biases we
refit the LDE model using the method of \citet[][STY]{sandage79} in
the formulation of \citet{efstathiou88}. The STY method fits the model
to the unbinned data, removing the biases discussed by
\citet{miyaji01}. We note, however, that it is not possible to include
the upper bounds on the bright-end of the LF in this formulation. For
this experiment we held $\alpha=-3.30$ fixed to the best-fit value of
the LDE model to the binned data. We also fixed the lowest redshift
characteristic density parameter to set a density scale, as the STY
method cannot determine the absolute normalization of the space
density. We find that all of the best fit parameters are within the
2$\sigma$ range of the values found for the binned LF. Moreover, the
shape of the evolution of $M_*(z)$ and $\phi_*(z)$ is relatively
unchanged from those obtained using the binned LFs (see
Fig. \ref{fg:evol}). The faint-end slope, $\beta$, is flatter than for
the binned LF by 2$\sigma$. While most of the $M_{*,k}$ parameters are
fainter than for the binned LF, this is due to the larger value of
$\beta$ and its covariance with $M_*(z)$. If we repeat the fits
holding $\beta=-1.30$ to match our previous fits, we find $M_{*,k}$
parameters in very good agreement with those obtained by fitting the
binned LF (see Table \ref{tab:qsofit_nohost}). Given that the changes
in the parameters are only modest, fitting the parametric models to
the binned luminosity functions is not an important source of
systematic errors and does not bias our conclusions.

\subsection{Comparison with Other Measurements}\label{ssec:qlf_comp}

In the previous sections, we constructed the QSO luminosity functions
of mid-IR and X-ray selected AGNs in the NDWFS Bo\"otes field from $0
\lesssim z \lesssim 6$ using several different functional forms to
describe the shape and evolution of the QLF. We settled on a standard
best-fit LDE model. In this section we make a broad comparison of our
results to those in the literature for optical, mid-IR and X-ray
selected AGNs. Detailed comparisons with other works can be found in
Appendix \ref{sec:qlf_comp_spec}. A significant complication to this
comparison is the very different wavelengths of our measurements
compared to other surveys. Wavelength corrections are particularly
important when comparing the predicted densities of quasars, as a
uniform magnitude limit must be assumed. We use the unreddened AGN
template presented in \citetalias{assef09} to convert between
wavelengths and make uniform comparisons between surveys. Some of the
QLFs we compare to, however, are measured at rest-frame UV, and hence
the conversion from our $J-$band measurements is very dependent on the
shape of the SED. As discussed in \citetalias{assef09}, the optical
slope of our template is bluer than typical AGN SEDs, as it tries to
match the bluest quasars in the sample rather than the average
object. To account for this, we also present some results using our
AGN template with $E(B-V)=0.05$, which better matches the UV part of
the \citet{richards06a} mean Type-I quasar SED. Table \ref{tab:colors}
shows the rest frame color $X - J$ used for each case discussed in
this Section, where $X$ is the corresponding rest-frame band used by
each study, for both values of the reddening. For reference, as
discussed in \citetalias{assef09}, we have assumed a reddening that
follows an SMC-like extinction curve for $\lambda < 3300$\AA\ and a
Galactic extinction curve at other wavelengths, with $R_V = 3.1$ for
both regimes.

\begin{deluxetable}{l l  c r c  c r c  c c}

\tablecaption{Color Transformations for QLF Comparison\label{tab:colors}}

\tablehead{QLF & Band X &\multicolumn{6}{c}{Color X--J} & Mag. System\\
               &        &  \multicolumn{3}{c}{\footnotesize{E(B--V)=0}}   &
  \multicolumn{3}{c}{\footnotesize{E(B--V)=0.05}}    &       }
\tabletypesize{\small}
\tablewidth{0pt}
\tablecolumns{12}

\startdata 

\citet{croom04}                & $b_j$    & & \phs1.30 & & & \phs1.45 & & Vega\\
\citet{croom09}                & $g(z=2)$ & &   --1.16 & & &   --0.50 & & AB\\
\citet{fan01}\tablenotemark{a} & 1450\AA  & & \phs0.05 & & & \phs0.76 & & AB\\
\citet{richards06b}            & $i(z=2)$ & &   --1.12 & & &   --0.78 & & AB\\ 
\citet{brown06}                & [8.0]    & &   --6.94 & & &   --6.98 & & Vega\\ 
\citet{hopkins07}              & $B$      & & \phs1.29 & & & \phs1.46 & & Vega\\ 

\enddata

\tablecomments{The color transformations used for the QLF comparisons
of \S\ref{ssec:qlf_comp} as derived from our AGN SED template for
E(B--V) values of 0 and 0.05. The colors are shown in the AB system
for QLF studies using bands with that calibration, while they are
shown in Vega otherwise.}

\tablenotetext{a}{\citet{wolf03} and \citet{jiang09} use the same band
as \citet{fan01}.}

\end{deluxetable}

\begin{figure}
  \begin{center}
    \plotone{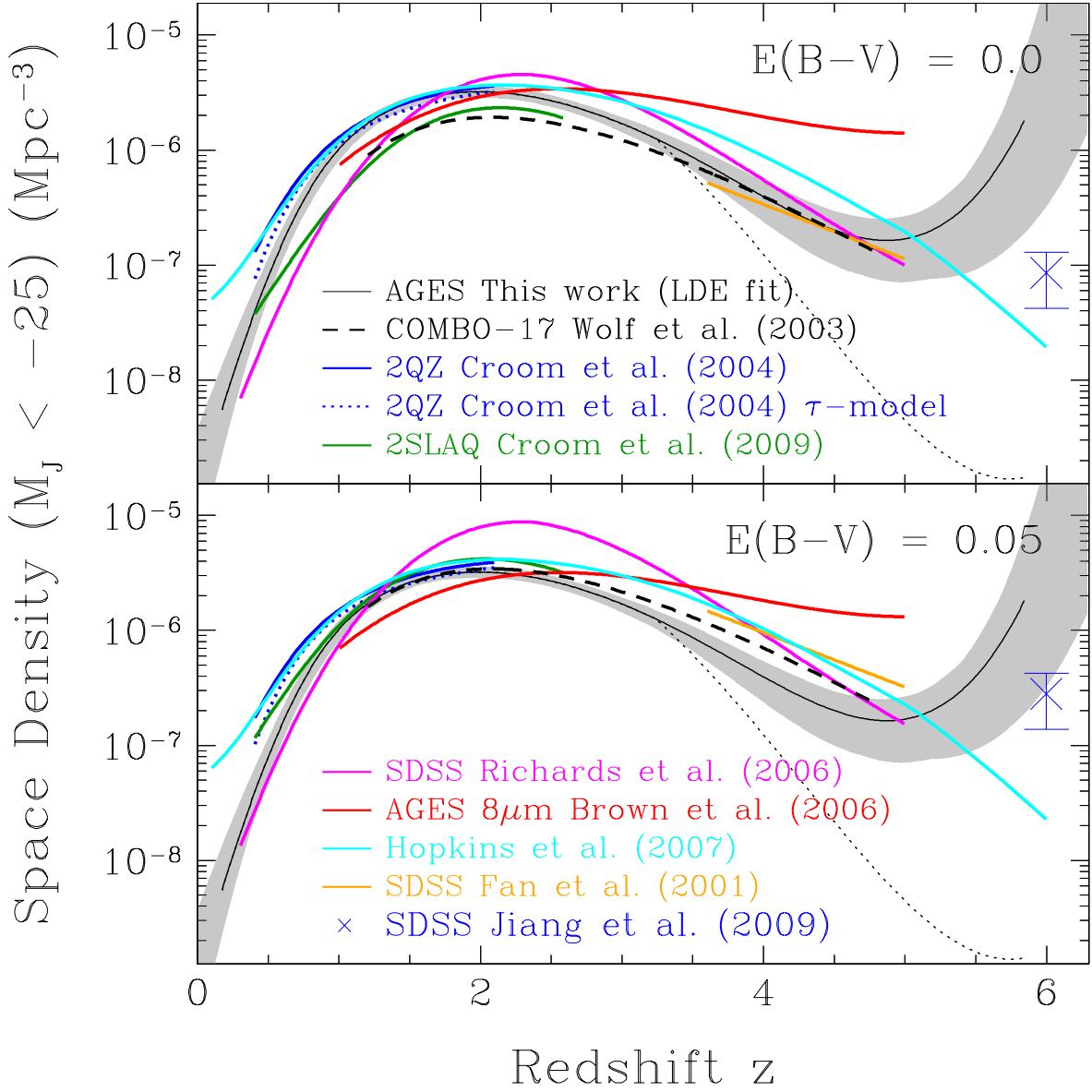}
    \caption{Space density of bright quasars as determined from our
    best-fit LDE model. The upwards turn at $z>5$ is unlikely to be
    real (see \S\ref{ssec:qlf_comp_spec_opthiz} for details). The gray
    shaded region shows the 2$\sigma$ ($\Delta \chi^2 \leq 4$)
    confidence region of our fit. We also show the space density
    predicted by the studies we discuss in \S\ref{ssec:qlf_comp}. We
    use our AGN template convert between the different bands of each
    study assuming no reddening ({\it{top}}) and $E(B-V)=0.05$
    ({\it{bottom}}). The dotted black line shows the space density
    predicted from the best-fit LDE model but assuming a bright limit
    for $M_{*,J}(z)$ of --28.}
    \label{fg:rho}
  \end{center}
\end{figure}

Figure \ref{fg:rho} shows the space density of bright AGNs ($M_J <
-25$) predicted by our best-fit LDE model as a function of redshift,
along with the predictions of the best-fit functional form of other
studies in the literature. In particular, we show the results of
\citet[2QZ survey]{croom04}, \citet[2SLAQ and SDSS]{croom09},
\citet[SDSS]{fan01}, \citet[SDSS]{jiang09}, \citet[AGES -- MIPS
  selected 8.0$\mu$m LF]{brown06}, \citet[COMBO--17]{wolf03} and
\citet{hopkins07}. A detailed description of each study (bands, target
selection and result comparison) is given in Appendix
\ref{sec:qlf_comp_spec}. Color transformations based upon our
unreddened AGN SED are appropriate for QLFs constructed in optical and
mid-IR rest-frame (2QZ, 8$\mu$m AGES and \citealt{hopkins07}), while
the reddened AGN template color conversion is more appropriate for UV
rest-frame QLFs (COMBO-17, 2SLAQ and SDSS). In general, Figure
\ref{fg:rho} shows the agreement is relatively good with most other
studies, although discrepancies are present. In most cases the
discrepancies can be attributed to the small sample sizes of some of
the studies \citep{wolf03,brown06,fan01} or of some of the redshift
ranges in our sample (again, see Appendix \ref{sec:qlf_comp_spec} for
a detailed comparison). Note the general agreement at the redshift of
the peak of our distribution ($z\sim 2$) with all other studies that
encompass this redshift. The largest disagreements comes from the
SDSS-based study of \citet{richards06b} at $z\sim 2$, however since
their sample is a sub-sample of that used by \citet{croom09}, which
shows remarkable agreement, the discrepancies are unlikely to be
important.

The agreement with the high redshift ($3.6 \leq z \leq 5.0$) results
of \citet{fan01} is very good in shape, although somewhat discrepant
in normalization. Because of their small sample size, \citet{fan01}
used a single power-law fit to the QLF, but found that the best-fit
slope ($\alpha = -2.58\pm 0.23$) was significantly flatter than at low
redshift, suggesting strong evolution was present in the bright-end
slope of the QLF. Because of this, the agreement with our results is,
at first glance, surprising, as we do not allow for evolution in the
slopes $\alpha$ and $\beta$, but can be explained by the fast
evolution we find for $M_{*,J}$. At $z>4$ our sample is not large
enough to accurately constrain $M_{*,J}$, but based on the $3 < z < 4$
redshift bin we can rule it out being fainter than $M_{*,J}\approx
-28$. Using the color conversion based our slightly reddened AGN SED
template to go from $J$ to the 1450\AA\ magnitude used by
\citet{fan01}, we find that the magnitude break is either within the
magnitude range of their sample or brighter than it (see
\S\ref{ssec:qlf_comp_spec_opthiz} for details). Thus, in our best-fit
LDE model, the QLF constructed by \citet{fan01} is in the transition
between the faint and bright ends, naturally explaining their best-fit
slope value being in between the low redshift measurements of $\alpha$
and $\beta$. We caution the reader, however, that a deeper and larger
sample at $z\sim4$, in which the QLF break is clearly visible, is
needed to eliminate the model dependence of this result and
corroborate our interpretation.

Figures \ref{fg:low_z_alpha_beta} and \ref{fg:high_z_alpha_beta} show,
for each study where we could make the comparison, the best fit
power-law indices $\alpha$ and $\beta$. The agreement between all
different studies (including ours) is generally bad, although we note
that our estimates for $\alpha$ and $\beta$ are generally within the
range determined by all other studies we show rather than at an
extreme.

X-ray surveys have found results that broadly agree with our
estimates. In particular, \citet{hasinger05} found that the soft X-ray
luminosity function (SXLF) of AGNs between $0 < z < 4.8$ is well
described by a double power-law functional form allowing for
luminosity dependent density evolution (LDDE). Recently,
\citet{aird09} found that the $2 - 10~\rm keV$ luminosity function of
AGNs between $0 < z < 3.5$ is equally well described by an LDDE model
and by a simpler combination of just luminosity and density evolution,
similar to our LDE model. While we cannot directly compare our results
to these X-ray luminosity functions because our AGN SED template does
not extend into this wavelength regime, the general trends agree in
that the evolution of the QLF must be in both luminosity and density.

\subsection{Comparison with the IRAC-selected QLF}

\begin{figure}
  \begin{center}
    \plotone{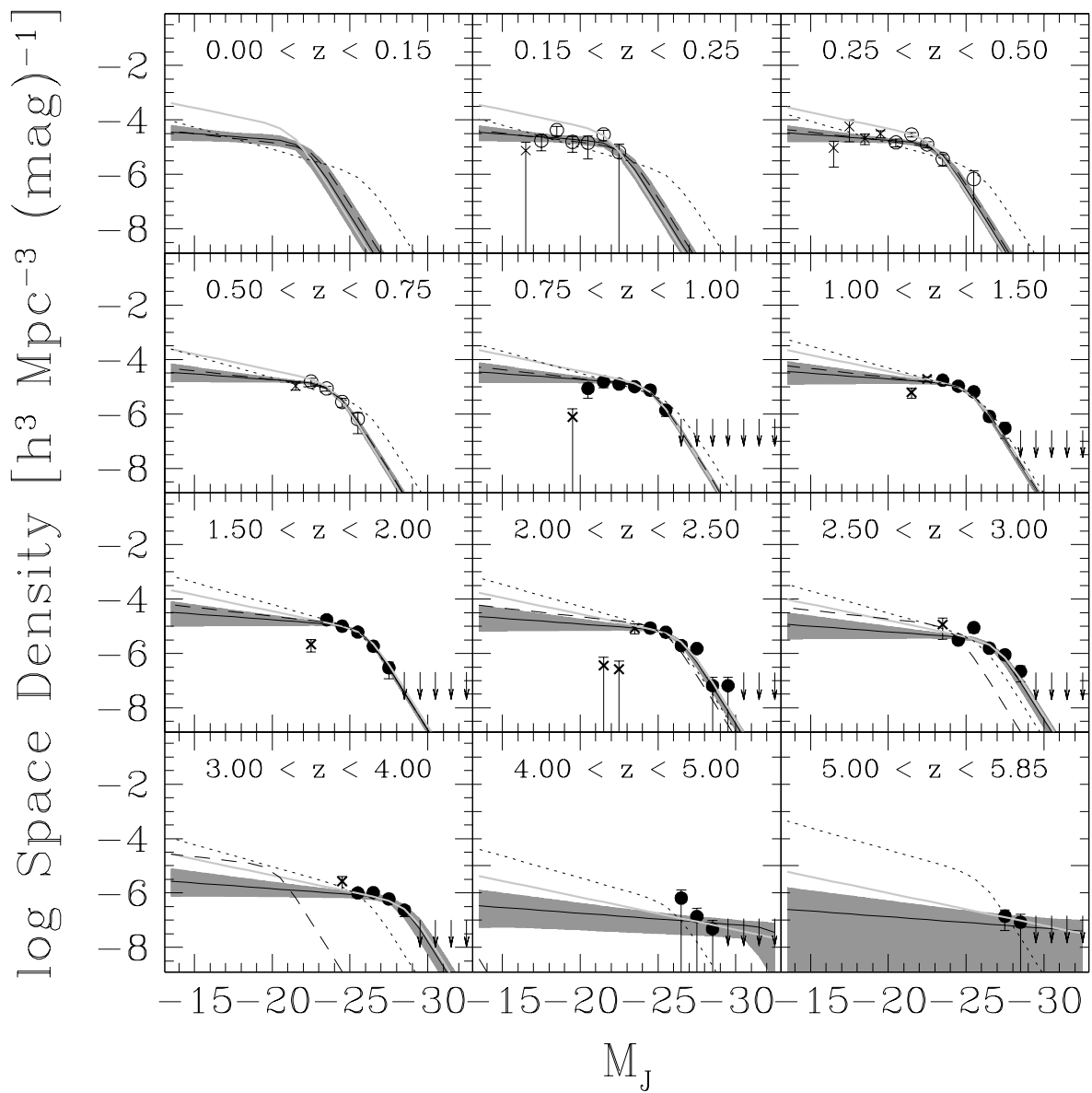}
    \caption{Same as Figure \ref{fg:no_host_lum_lfs} but constructed
    only from IRAC selected sources. For reference, the solid light
    gray line shows the best-fit LDE model to the full sample (see
    Fig. \ref{fg:no_host_lum_lfs}).}
    \label{fg:irac_only_no_host_lum_lfs}
  \end{center}
\end{figure}

In \S\ref{ssec:IRAC_QSO_selection} we argue that the selection
criteria used to target AGNs by their IRAC colors, which consists of
slight modifications of the \citet{stern05} criterion, are subject to
two strong biases, namely losing AGNs that are faint in comparison to
their host galaxies and in the redshift range $4 < z < 5$. In this
section, we compare the AGN luminosity function derived in the
previous sections and that constructed by considering only the IRAC
selected objects. Figure \ref{fg:irac_only_no_host_lum_lfs} shows the
host-corrected IRAC-only QLF along with its corresponding best-fit
LDE, PLE and PDE models. The best-fit models are also summarized in
Table \ref{tab:irac_qsofit_nohost}. Note that there are no
IRAC-selected AGNs in the lowest redshift range ($z<0.15$).

\begin{deluxetable}{l r r r}

\tablecaption{IRAC Selected Host Corrected J-band QSO Luminosity Function Parametric Fits \label{tab:irac_qsofit_nohost}}
\tablehead{Parameter & LDE & PLE & PDE}
\tabletypesize{\small}
\tablewidth{0pt}
\tablecolumns{4}

\startdata 

$\chi^2$                               &             57       &            143       &            154      \\
$\chi^2_{\nu}$                         &          1.427       &          3.254       &          3.571      \\
$\alpha$                               &  --3.29 $\pm$   0.05 &  --3.29 $\pm$   0.05 &  --3.29 $\pm$   0.05\\
$\beta$                                &  --1.11 $\pm$   0.07 &  --1.16 $\pm$   0.06 &  --1.40 $\pm$   0.04\\
$M_{*,J}     $                         &         \nodata      &         \nodata      & --26.03 $\pm$   0.15\\
$M_{*,J}(z_1)$                         & --23.56 $\pm$   0.21 & --23.81 $\pm$   0.18 &         \nodata     \\
$M_{*,J}(z_2)$                         & --24.84 $\pm$   0.14 & --24.99 $\pm$   0.14 &         \nodata     \\
$M_{*,J}(z_3)$                         & --26.07 $\pm$   0.13 & --25.80 $\pm$   0.14 &         \nodata     \\
$M_{*,J}(z_4)$                         & --30.11 $\pm$   0.77 & --15.74 $\pm$   2.01 &         \nodata     \\
$\rm{log}\ \phi_*(0)$\tablenotemark{a} &         \nodata      &  --4.97 $\pm$   0.05 &         \nodata     \\
$\rm{log}\ \phi_*(z_1)$                &  --4.81 $\pm$   0.12 &         \nodata      &  --5.91 $\pm$   0.13\\
$\rm{log}\ \phi_*(z_2)$                &  --4.86 $\pm$   0.08 &         \nodata      &  --5.71 $\pm$   0.10\\
$\rm{log}\ \phi_*(z_3)$                &  --4.89 $\pm$   0.06 &         \nodata      &  --5.38 $\pm$   0.07\\
$\rm{log}\ \phi_*(z_4)$                &  --5.04 $\pm$   0.06 &         \nodata      &  --5.16 $\pm$   0.06\\
$\rm{log}\ \phi_*(z_5)$                &  --6.72 $\pm$   0.15 &         \nodata      &  --6.28 $\pm$   0.17\\
$M_{*,J}(0)$                           &      --21.52         &      --22.38         &         \nodata     \\
$k_1$                                  &         2.01         &         1.20         &         \nodata     \\
$k_2$                                  &       --0.81         &       --0.05         &         \nodata     \\
$k_3$                                  &         0.13         &       --0.10         &         \nodata     \\

\enddata

\tablecomments{The best fit parameters of each of our three models for
the QLF with the formal 1$\sigma$ error-bars from the Levenberg --
Marquardt fitting algorithm. For an easier comparison with other
results in literature, we also show the values of the parameters of
the more commonly used functional form for the evolution of $M_*$
shown in equation (\ref{eq:mstar_others}). We show no error-bars for
this values as correlations between them can be highly significant.}

\tablenotetext{a}{~$\rm h^3\ \rm Mpc^{-3}\ \rm mag^{-1}$}

\end{deluxetable}

In general, the IRAC-selected QLF has a significantly flatter faint
end slope $\beta$ than that of our full sample. This is consistent
with Figure \ref{fg:Lagn_Lhost}, where we show that intrinsically
fainter AGN typically have larger $\rm L_{\rm Host}/ \rm L_{\rm AGN}$
ratios because the method is only sensitive to objects whose mid-IR
colors are dominated by the AGN rather than by the host. Although the
minimum $L_{\rm AGN} / L_{\rm Host}$ ratio for selection is dependent
on the source's redshift, the AGN's reddening and the spectral type of
the host, it roughly corresponds to requiring $L_{\rm AGN}/L_{\rm
Host}\gtrsim 1$. Since lower $L_{\rm AGN}$ sources tend to have lower
values of $L_{\rm AGN} / L_{\rm Host}$, the IRAC selection criteria
eliminates more faint than bright AGNs from our sample. Few faint AGNs
are selected by their IRAC colors, even though the region is well
populated by the MIPS and X-ray selected AGNs.

Figure \ref{fg:evol} shows the evolution of $M_{*,J}(z)$ and
$\log[\phi_*(z)]$ along with those of our full sample. The evolution
of the $M_{*,J}$ parameter is largely unaffected by the biases of the
IRAC AGN selection, but the evolution of the $\log[\phi_*(z)]$ is
qualitatively different at low redshifts ($z<1$). While for the full
sample $\log[\phi_*(z)]$ increases strongly with decreasing redshift,
for the IRAC selected sample it is flat. This is consistent with our
discussion of the flatter faint end slope. Although the effects of the
IRAC selection biases have a stronger effect upon lower luminosity
objects, at $z<1$ they still affect the space density estimate for
AGNs of magnitude $M_{*,J}(z)$ with some significance. However, since
$M_{*,J}(z)$ becomes brighter with increasing redshift, at $z>1$ the
IRAC selection biases have little effect on the estimated space
density of objects brighter than the characteristic magnitude
$M_{*,J}(z)$.

\section{Conclusions}

We derived the $J-$band QSO luminosity function both with and without
the flux contribution of the host galaxy, using 1838 mid-IR and X-ray
selected AGN in the redshift range $0 < z < 5.85$. Our sample is based
upon the multi-wavelength photometry of the NDWFS survey and the
spectroscopic follow up of IRAC selected objects by AGES. We used the
set of low resolution SED templates for galaxies and AGNs presented in
a previous paper \citepalias{assef09} to accurately characterize the
properties of the selection function and take them into account when
building the QSO luminosity function (QLF). The IRAC color selection
of sources by the AGES survey used variations of the \citet{stern05}
selection criteria. Based on our AGN SED template, such selection
criteria are biased towards high Eddington ratio objects so that
mid-IR colors are dominated by the AGN, and against redshifts near
$4\lesssim z \lesssim 5$ where the mid-IR color tracks make a ``blue
loop'' outside the color selection region due to H$\alpha$ emission
redshifted into the [3.6] IRAC channel. Our IRAC-selected sample
contains few objects in this redshift range (see Figure
\ref{fg:n_agn_z}) relative to both higher and lower
redshifts. Including MIPS and X-ray selected sources largely solves
these problems.

Using our low resolution templates we have corrected the objects for
host contamination and have also estimated the bolometric QLF of our
sources. Our host corrected luminosity functions show that the space
density of bright AGN strongly decreases from $z=3$ to $z=0$. The
best-fit model to the LF suggests that, in contrast, the space density
of faint quasars is relatively constant in this redshift range. While
the data is consistent with this behavior, it is not strongly
constraining. The redshift at which this steep decline begins is also
a strong function of luminosity. At $z>3$ we observe a decrease in the
space density of AGNs at all luminosities.

We have modeled the QLF by a double power-law with redshift
independent slopes. Because of our modest sample volume, the
bright-end slope ($\alpha$) cannot be well constrained, and we imposed
a prior based on the value measured by \citet{croom04} with the 2QZ
survey. Pure luminosity (PLE) and pure density evolution (PDE) fail to
describe the evolution of the QLF over the whole redshift range of our
sample. A combination of both is needed to explain the evolution of
the QLF with cosmic time. Our combined evolution model (LDE) yields a
significantly better fit than either the PLE or PDE models. This is in
agreement with measurements of the QLF by \citet{croom09} and of the
X-ray luminosity function (XLF) by \citet{aird09}. Our results agree
relatively well with other measurements at all redshifts. In
particular, we find good agreement with the results of low and high
redshift optically selected QLFs and with other measurements that link
both redshift ranges. While our fits do not allow for an evolving
bright-end slope, we still find good agreement with the $3 < z < 5$
measurements of \citet{fan01} and provide a tentative explanation for
the flattening of the bright-end slope they find, although we note
that a deeper sample is needed to corroborate it. Because our sample
does not suffer from the incompleteness of optical selection methods
at $2.5\lesssim z \lesssim 3.5$ it provides an accurate link between
low and high redshift QLFs despite the comparatively lower number of
objects used.

Finally, we have estimated the QLF using only the IRAC selected AGNs
in order to understand the impact of the IRAC selection biases upon
the best fit QLF. We find a significantly flatter faint end slope
$\beta$ in comparison to our full sample and a flatter evolution of
$\phi_*$ at $z<1$. However, at $z>1$ we find that $\phi_*$ is
generally in good agreement with the estimates obtained using the full
sample, and that $M_{*,J}$ is not significantly affected by the biases
of the IRAC selection in the full redshift range of our sample.

\acknowledgements 

We would like to thank all the people in the NDWFS, FLAMEX and SDWFS
collaborations that did not directly participate in this work. We also
thank the anonymous referee for comments and suggestions that helped
improve this work. Support for MB was provided by the W. M. Keck
Foundation. The work of DS was carried out at Jet Propulsion
Laboratory, California Institute of Technology, under a contract with
NASA. The AGES observations were obtained at the MMT Observatory, a
joint facility of the Smithsonian Institution and the University of
Arizona. This work made use of images and/or data products provided by
the NOAO Deep Wide-Field Survey \citep{ndwfs99,jannuzi05,dey05}, which
is supported by the National Optical Astronomy Observatory
(NOAO). This research draws upon data provided by Dr. Buell Jannuzi
and Dr. Arjun Dey as distributed by the NOAO Science Archive. NOAO is
operated by AURA, Inc., under a cooperative agreement with the
National Science Foundation.

\appendix

\section{Detection probability of X-ray sources}\label{sec:xray_det}

In \S\ref{ssec:qso_lf_det} we discussed our implementation of the
\citet{page2000} method for estimating $V/V_{\rm max}$ (see
eqn. [\ref{eq:v_vmax}]). This implementation depends upon the
probability that a given object, at a certain redshift $z$ and with a
certain magnitude $M_J$, is included in the sample. For the IRAC and
MIPS selection criteria, the detection probability is either 0 or 1,
depending on whether the object passes the different color and
magnitude cuts. This implicitly assumes that the errors in the
measurements are fractionally small, which is sensible for the UV to
mid-IR photometry, as the photometric catalogs from which they were
selected are typically much deeper than the spectroscopic sample we
use. This is not true for the X-ray selected sample. Most sources in
the XBo\"otes survey are detected with only a few counts, and hence
the full Poisson probability distribution of their errors must be
considered.

To determine the X-ray detection probability of our sources as a
function of redshift and $J-$band absolute magnitude, we use the
following procedure. Consider an object located at redshift $z_0$ with
$J-$band absolute magnitude $M_{J,0}$ and detected with $N_0$ X-ray
counts. The probability that we could have detected it with $N_D$
counts if located at an arbitrary redshift $z$ with an arbitrary
absolute magnitude $M_J$ can be expressed as
\begin{equation}
P(N_D|N_0,z,M_J) = \int_0^{\infty} P[N_D|N(N_T,z,M_J)]\ P(N_T|N_0)\ dN_T,
\end{equation}
\noindent where $N_T$ is the source's ``true'' number of X-ray counts
at redshift $z_0$ and absolute magnitude $M_{J,0}$ and $N(N_T,z,M_J)$
is the expected number of counts at redshift $z$ and absolute
magnitude $M_J$. This can be re-written as
\begin{equation}\label{eq:prob}
P(N_D|N_0,z,M_J) \propto \int_0^{\infty} P[N_D|N(N_T,z,M_J)]\
P(N_0|N_T)\ P(N_T)\ dN_T,
\end{equation}
and now both terms, $P[N_D|N(N_T,z,M_J)])$ and $P(N_0|N_T)$, are given
by Poisson distributions. The probability of observing objects with
$N_T$ X-ray counts is proportional to the number of sources with this
intrinsic number of counts. \citet{kenter05} determined for the
XBo\"otes survey that the number of sources per unit flux, $n_S$, is
well described by a broken power-law of the shape
\begin{equation}\label{eq:p_nt}
n_S\ \propto\ \left\{ \begin{array}{rl}
    N_T^{-1.74}&  N_T \leq N_B\\
    N_T^{-2.60}&  N_T > N_B
    \end{array} \right. ,
\end{equation}
\noindent where $N_B \approx 7$ is the number of counts of the break
in the power law distribution. However, \citet{kenter05} notes that
the flux of the break is very poorly constrained and the 1$\sigma$
interval of the best-fit value puts it between 3 and 25 counts. Since
for large values of $N_0$ the probability is little affected by the
value of the power-law index, we assume that the number of sources per
unit flux is a single power-law $n_S \propto N_T ^{-y}$, with $y =
1.74$, corresponding to the faint end of equation \ref{eq:p_nt}.

Assuming that all sources have an X-ray spectrum well described by a
power law with photon index $\Gamma = 1.7$, and hence $F_{\nu} \propto
\nu^{-\Gamma+1}$, it can be shown that $N(N_T,z,M_J) = f\ N_T$, where
\begin{equation}
f\ =\ \left(\frac{1+z}{1+z_0}\right)^{2-\Gamma}\
\left[\frac{D_L(z_0)}{D_L(z)}\right]^2\ 10^{-0.4(M_J-M_{J,0})}.
\end{equation}
Substituting in equation \ref{eq:prob} and calculating the integral,
we find that the probability of observing a source seen to have $N_0$
counts at redshift $z_0$ and magnitude $M_{J,0}$ with $N_D$ counts at
redshift $z$ and magnitude $M_J$ is
\begin{equation}
P(N_D|N_0,z,M_J) \propto f^{N_D}\ (1+f)^{y-N_D-N_0-1}\
\frac{\Gamma(1+N_D+N_0-y)}{\Gamma(1+N_D) \Gamma(1+N_0)} ,
\end{equation}
\noindent where $\Gamma(x)$ corresponds to the $\Gamma$
function. Finally, the probability of a given object detected with
$N_0$ counts at $z = z_0$ and magnitude $M_J$ producing 4 or more
counts when located at redshift $z$ and with $J-$band absolute
magnitude $M_J$ is given by
\begin{equation}
P_{det}\ =\ \sum^{\infty}_{N_D=4}\ c(N_D)\ P(N_D|N_0,z,M_J)\ \left/\
\sum_{N_D=0}^{\infty}\ P(N_D|N_0,z,M_J)\right. ,
\end{equation}
\noindent where $c(N_D)$ is the completeness of sources with $N_D$
counts in XBo\"otes as determined by \citet{kenter05}.

Note that we have assumed here that the X-ray flux scales linearly
with changes in the $J-$band flux for each object. The X-ray to
optical flux ratio ($\alpha_{\rm ox}$) of AGNs is known to scale with
their rest-frame, unreddened, UV flux, but this dependence is weak
\citep{strateva05}. However, since the changes in $M_J$ for estimating
$V/V_{\rm max}$ are small ($\lesssim 1~\rm mag$), assuming strict
linearity for scaling the X-ray fluxes instead of a more complex
relation does not affect our results.

\section{Detailed Comparison with Other QLF Studies}\label{sec:qlf_comp_spec}

In this section we present a detailed comparison of our results to
those of other studies discussed broadly in \S\ref{ssec:qlf_comp}. For
simplicity, we have divided the different studies according to the
properties of their samples into those based on optically selected low
redshift samples \citep{croom04,croom09}, based on optically selected
high redshift samples \citep{fan01,jiang09}, those with samples
spanning broad redshift ranges
\citep{wolf03,richards06b,brown06,hopkins07} and those based on X-ray
observations \citep{hasinger05,aird09}.

\subsection{Optically Selected, Low Redshift QLFs}

\citet{croom04} combined the observations of the 2dF QSO Redshift
Survey (2QZ) and the 6dF QSO Redshift Survey (6QZ) to study the
optical ($b_j$ band) QSO luminosity function in the redshift range
between 0.4 and 2.1 up to a magnitude limit of $b_j = 20.85$. This
corresponds to $I = 20.47$ for our AGN template with no reddening. The
low redshift limit of 0.4 was chosen to avoid dealing with extended
sources, while the upper limit of 2.1 was chosen because of selection
incompleteness caused by stars and the limits of the UVX targeting
method employed \citep{smith05}. Their full sample consists of 21222
QSOs, most of which are brighter than the break magnitude because of
the bright limiting magnitudes. This means that their measurement of
the bright-end slope is very precise, while their constraints for the
faint end are weak. They parametrize the QLF using the double
power-law of equation (\ref{eq:phi_mod}) and assume pure luminosity
evolution described using either an exponential scaling of $L_{*}$
with look-back time defined by an e-folding time (the $\tau$-model),
or by a second order polynomial (eqn. [\ref{eq:mstar_others}]). While
both fits to their data are acceptable, the $\tau$-model is
better. Figure \ref{fg:rho} shows, for both models, the space density
of bright quasars ($M_J < -26$) as a function of redshift. Both of
their parametrizations predict a space density of bright quasars in
agreement with our best-fit models at $z\simeq 2$, close to the peak
in bright quasar activity. Below $z\simeq 1$, however, they
significantly overpredict the number of bright quasars compared to our
estimate. Our sample contains few objects with $M_J < -26$ and $z<1$
(see Figure \ref{fg:no_host_lum_lfs}) and so this difference may not
be highly significant. By construction, our values of $\alpha$ agree
very well with those of their polynomial fit (see
\S\ref{ssec:qso_lf_det}), while our values of $\beta$ disagree at the
$\sim 2 - 3\sigma$ level.

\citet{croom09} studied the QLF in the $M_g(z=2)$ band (absolute
rest-frame $g-$band $K-$corrected to $z=2$) between $0.4 < z < 2.6$
using spectroscopic observations of optically selected QSOS from the
2SLAQ survey combined with the low redshift part of the SDSS QSO
sample of \citet[discussed later in this section]{richards06b}. Their
sample contains 15073 QSOs, is limited to objects with $g<21.85$ and is
corrected for host galaxy contamination using statistical relations
between the host and quasar luminosities and redshift. They find
strong signs of down-sizing and determine that the best-fit functional
form for their sample is given by a combined luminosity and density
evolution model, similar to our LDE fit, but allowing $\alpha$ to
evolve and limiting the $M_{*,g}(z)$ and $\log[\phi_*(z)]$ polynomials
to second order in $z$ (see eqns. [\ref{eq:mstar_z}] and
[\ref{eq:phi_star_z}]). Their sample cannot be well fit by simple PLE
or PDE models, in agreement with our results (see
\S\ref{ssec:qlf_fit}). Figure \ref{fg:low_z_alpha_beta} shows their
best-fit values of $\alpha$, as a function of redshift, and
$\beta$. Between $1<z<2$, their $\alpha$ agrees relatively well
($<3\sigma$) with our best-fit LDE parameter, but at higher and
lower redshift it is significantly different. On the other hand,
$\beta$ agrees very well with our LDE estimate. For completeness, we
show the values of $\alpha$ and $\beta$ estimated by
\citet{richards05} using an earlier version of the 2SLAQ data
set. \citet{richards05} assumed the same model and best-fit parameter
values of \citet{croom04} and used the 2SLAQ QLFs to only fit $\beta$,
which is the only unreliable value of the 2QZ survey results. The
estimate of $\beta$ by \citet{richards05} agrees well with ours (the
combination of both parameters is $\sim 1.5\sigma$ different than our
estimates). In Figure \ref{fg:rho} we show the predicted space density
of bright quasars, and when considering the slightly reddened AGN
templates, the agreement is very good over the whole redshift range.

\subsection{Optically Selected, High Redshift QLFs}\label{ssec:qlf_comp_spec_opthiz}

High redshift quasars ($z>3$) are found by different color selection
methods and are generally fit as a separate population. \citet{fan01}
used a sample of 39 SDSS QSOs with $i<20$ selected by their observed
optical colors and spanning $3.6 < z < 5.0$ to study the rest-frame
1450\AA\ QLF. Because their sample is not very deep and contains few
objects, they model the QLF by a single power-law with pure luminosity
evolution. They divided their sample into three redshift bins with
edges of 3.3, 3.9, 4.4 and 5.0, and either two or three luminosity
bins at each redshift. Their results show a bright-end slope of
$-2.58\pm 0.23$ (see Figure \ref{fg:high_z_alpha_beta}), which is
considerably shallower than the value of $\approx -3.31$ found by
\citet{croom04} using the 2QZ survey for low redshifts. This argues
for a strong evolution of the bright-end slope of the QLF. Nonetheless
our best-fit LDE model is not inconsistent with the results of
\citet{fan01}, despite the fact that we do not allow for evolution in
$\alpha$ or $\beta$. The evolution of $M_{*,J}$ we find is fast enough
to suggest that the magnitude bins spanned by their sample likely fall
in the transition region between the bright and faint ends of the
QLF. If we use our AGN template (with a reddening of 0.05 mags) to
transform our estimates of $M_{*,J}$ into $M_{*,1450}$ at each of
their redshift bins, we obtain $M_{*,1450} = (-28.19, -30.36, -34.88)$
for $z = (3.75, 4.15, 4.7)$, all fainter than the objects used in
their sample (see Figure 1 of \citealt{fan01}), arguing that their
objects are in the faint end of the QLF. Note, however, that our
constraints on $M_*$ at $z>4$ are weak, although the absence of very
bright quasars leads to strong lower bounds on $M_{*,J}$. At $z=3.75$
our data rules out an $M_{*,J}$ significantly fainter than $-28$ (see
Fig. \ref{fg:no_host_lum_lfs}) which corresponds to
$M_{*,1450}=-26.35$ and falls within the magnitude range of the
\citet{fan01} sample ($-27.5 < M_{1450} < -25.5$). In other words, the
fast evolution in $M_*$ suggested by our sample gives a natural
explanation to the observed flattening of $\alpha$. Figure
\ref{fg:rho} shows the predicted space density of bright quasars from
the study of \citet{fan01}. The agreement is good over their whole
redshift range when using the unreddened templates to convert the
$M_J$ limit into 1450\AA, but somewhat discrepant in normalization,
but not shape, for the reddened template. This discrepancy is probably
not very significant due to the small number of objects that comprise
their sample.

\citet{jiang09} studied the high redshift 1450\AA\ QLF from a combined
sample of 27 quasars at $z\sim 6$ with a limiting magnitude of 21.8 in
$z-$band. The QLF of their sample is well described by a single
power-law with index $-2.6\pm 0.3$, shown in Figure
\ref{fg:high_z_alpha_beta}, consistent with the redshift $z\sim 3$ QLF
of \citet{fan01}. The $M_{1450}$ absolute magnitude range of their
sample ($-28 \lesssim M_{1450} \lesssim -25$) is similar to that of
the \citet{fan01} sample, and hence the same caveats with regards to
the change in the bright-end slope at high redshifts apply. In Figure
\ref{fg:rho} we show the predicted space density compared to that of
our favored fitting model. Our results significantly overpredict the
number of bright quasars as compared to \citet{jiang09}, however this
may be due simply to our inability to constrain $M_{*,J}$ at $z\sim
6$ (see discussion in previous paragraph). The same Figure shows the
predicted space density if we assume a bright limit of $M_{*,J}=-28$
for our fits, and in that case we severely underestimate the space
density of bright quasars in comparison to \citet{jiang09}. Since the
real case likely lies between this lower bound and the value
estimated from our fits, this discrepancy is unlikely to be
significant.

\subsection{Broad Redshift Range QLFs}

There have been several earlier attempts to construct the QLF on both
sides of the peak density at $z\sim 2.1$. \citet{wolf03} studied the
rest-frame UV (1450\AA) luminosity function of 192 quasars from the
COMBO-17 survey (0.78 deg$^2$) in the redshift range $1.2 < z <
4.5$. This survey used a combination of 17 narrow, medium and broad
bands to securely identify quasars over this redshift range and
determine photometric redshifts. The lower redshift limit was chosen
to avoid dealing with host galaxy contamination to the SED, while the
higher limit was chosen to eliminate 3 bright QSOs whose existence was
grossly inconsistent with their lower redshift predictions and
predictions from other studies. Their sample was equally well
described by pure luminosity and pure density evolution and they did
not use a combined evolution model. The parametric form chosen by
\citet{wolf03} is a polynomial in $\log \phi$, and so we cannot
directly compare model parameters. However, we can compare the
expected space density of bright quasars, which is shown in Figure
\ref{fg:rho} for their pure density evolution model. The predicted
redshift for the peak of the bright quasars density is $z_{peak} =
2.08$, consistent with the value we find of $z_{peak} = 2.0\pm0.1$
($2\sigma$ error bars). The agreement is good at all redshifts, given
the small number of quasars in the sample of \citet{wolf03},
regardless of the reddening value assumed for our AGN template.

\citet{richards06b} also measured the QLF evolution from low to high
redshift. Using a sample of 15343 SDSS QSOs detected in 1622 deg$^2$
to measure the redshift 2 $i$-band ($i$-band K-corrected to $z=2$) QSO
luminosity function between $0.3 < z < 5.0$ up to a limiting observed
magnitude of $i = 19.1$ for $z<3$ and $i=20.2$ for $z>5$. These limits
correspond to $I<18.6$ and $I<19.7$ respectively for our unreddened
AGN template. Their sample was selected by optical colors and by radio
emission from the FIRST survey. Because of the optical color
selection, their sample is very incomplete at redshifts between $\sim
2.5$ and $\sim 3$, although this is partially alleviated by the
inclusion of the radio sources. Because of the shallowness of their
survey, \citet{richards06b} chose to model the QLF as a single
power-law. They assume a pure luminosity evolution model since a
single power-law description of the QLF cannot separate the effects of
density and luminosity evolution. They also allow for the bright-end
power-law index to evolve after redshift $2.4$, as it is strongly
favored by their data. As shown in Figure \ref{fg:rho}, our prediction
for the number of bright quasars as a function of redshift agrees
generally well with that of \citet{richards06b} at low redshift when
the reddened AGN template is used to convert between magnitudes. At
$z\sim 2$, however, the agreement is relatively poor, significantly
higher than our prediction. However, because their sample is a subset
of the \citet{croom09} sample, and their predictions agree very well
with ours in this redshift range, this is unlikely to be very
significant. At higher redshift the differences in predictions
decrease significantly and agree very well at $z\sim 5$.

\citet{brown06} studied the [8.0] luminosity function using a sample
of 183 MIPS 24$\mu$m selected QSOs brighter than 1mJy with redshifts
between 1 and 5 taken from an earlier release of the AGES
survey. \citet{brown06} did not include objects with $z<1$ in order to
avoid complications with the spectroscopic completeness for extended
sources. They fit their luminosity function with a single power-law
and assumed pure luminosity evolution using the same redshift
polynomial we considered here (although in the form of
eqn. [\ref{eq:mstar_others}]). They find a value of $\alpha = -2.75$,
much shallower than our bright slope. This is not unexpected, as their
sample has sources from both sides of the magnitude break, but it is
too small to fit $\alpha$ and $\beta$ independently. Figure
\ref{fg:rho} shows the evolution of the space density of bright
quasars determined by \citet{brown06}. The space density we determine
peaks at $z_{\rm peak} = 2.0\pm0.1$, well below the peak at $z_{\rm
peak} = 2.6\pm 0.3$ found by \citet{brown06} although only 2$-\sigma$
discrepant. The predicted number of bright quasars from their best-fit
functional form roughly agrees with that of our best-fit model at
$z\lesssim 3$, but the agreement is more discrepant at higher
redshifts. The discrepancies are, however, not very significant due to
the small number of objects that went into building this sample.

In this context, it is worth noting the work of \citet{hopkins07}, who
studied the evolution of the bolometric QLF from $z\sim 0$ and $z\sim
6$ from a compilation of observed binned QLFs from many surveys in the
literature across many wavelengths, from X-ray to 15$\mu$m (see their
Table 1 for a list of all surveys used). A detailed treatment of
reddening and the AGN SED was done in order to combine all the samples
accordingly, although the required assumptions may introduce
systematic errors into their final results. \citet{hopkins07} found
the resulting bolometric QLF was best described by a double power-law
(eqn. [\ref{eq:phi_mod}]) with evolution in the characteristic
bolometric luminosity (of a different form than that we assume), and
in the slopes $\alpha$ and $\beta$. Figure \ref{fg:high_z_alpha_beta}
shows the values of the latter two parameters as a function of
redshift. At $z=0$, both $\alpha$ and $\beta$ are highly discrepant
with most other measurements. The value of $\beta$ at low redshifts is
much lower than our LDE fit and all other literature measurements we
present here, but agrees well with the value from our PLE and PDE
fits. However, $\alpha$ is still inconsistent at the 3$\sigma$ level
from them. At $z=1$, the combination of both parameters coincide at
the $3\sigma$ level well with those determined from our LDE fit, where
$\alpha$ is more similar than $\beta$.  At higher $z$, the value of
$\beta$ is roughly consistent with that of our LDE fit, but the value
of $\alpha$ is lower by a very significant amount. Figure \ref{fg:rho}
shows the prediction of their best-fit functional form for the density
of bright quasars. We have incorporated their results into this plot
by using the C routine \verb+qlf_calculator+ provided by
\citet{hopkins07} to generate B-band luminosity functions. The
prediction is generally above that of our best-fit functional form,
but seems to agree at $z\sim2$. Their best-fit model predicts a
density peak at $z_{\rm peak} = 2.14$, consistent with our estimate of
$z_{\rm peak} = 2.0\pm0.1$ at about 3$\sigma$. At lower $z$, their
functional form overpredicts the number of bright QSOs relative to our
results but is consistent with the results of \citet{croom04}. This is
not highly surprising, as the 2QZ survey sample is a subset of the
data used by \citet{hopkins07}.

\subsection{X-ray QLFs}

X-ray surveys have found results that broadly agree with our
measurements of the QLF, although we cannot directly compare them with
our results as our templates do not extend into the X-ray regime. In
particular, \citet{hasinger05} studied the soft X-ray luminosity
function (SXLF) between $0 < z < 4.8$ with a combined sample from
several ROSAT, XMM-{\it{Newton}} and {\it{Chandra}} surveys. The
different samples are limited to relatively bright X-ray fluxes in
order to minimize incompleteness in the redshift measurements. Because
of this, their sample does not provide very good constraints on the
faint end of the SXLF. \citet{hasinger05} found that the evolution of
the SXLF is not well described by pure luminosity evolution, but is
well fit by a luminosity-dependent density evolution (LDDE)
model. This LDDE model is not similar to any of the models we discuss
in our sample, however \citet{croom09} finds that their 2SLAQ and SDSS
combined optical sample is better described by a combination of
luminosity and density evolution rather than by an LDDE model. The
analysis of \citet{hasinger05} shows that the evolution of low
luminosity AGNs is qualitatively different than that of brighter AGNs,
similar to our results (see \S\ref{ssec:qlf_fit} and Figure
\ref{fg:no_host_z_lfs}). They find that the space density of AGNs
peaks at lower redshifts for fainter objects than for brighter ones,
but also that it starts declining at lower redshifts for fainter AGNs,
unlike what we see in Figure \ref{fg:no_host_z_lfs}.

Recently, \citet{aird09} studied the 2 -- 10 keV X-ray luminosity
function (XLF) between redshifts $0 < z < 3.5$ from a combined sample
of the {\it{Chandra}} Deep Field North and South surveys, the AEGIS-X
survey, the {\it{ASCA}} Large Sky and Medium Sensitivity surveys and
other smaller surveys. To minimize the problems of optical
spectroscopic identification of the X-ray sources, they used a
combination of spectroscopic and photometric redshifts. Their sample
is composed of low redshift hard X-ray selected objects (limited to
$z<1.2$ to minimize the effects of catastrophic photometric redshift
errors) and of high redshift objects ($z\gtrsim 2$) selected by a
combination of soft X-ray detection and optical colors. To build the
XLF, they used a Bayesian approach to account for errors in the X-ray
flux measurements and photometric redshifts of their sources, as well
as the Eddington bias, in combination with detailed modeling of their
selection functions. \citet{aird09} find that their sample is equally
well described by an LDDE model and by a combination of luminosity and
density evolution, similar to our LDE model but with a different
parametrization. They also find that the space density of faint AGNs
evolves differently than that of brighter objects, however the
redshift of the peak of AGN activity shifts much more weakly with
luminosity than found by \citet{hasinger05} and other XLF
studies. They also find that the decline to higher redshifts is much
weaker than those found by other XLF studies.


\begin{thebibliography}{99}

\bibitem[Aird et al.(2010)]{aird09} 
  Aird, J., et al.\ 2010, \mnras, 401, 2531 

\bibitem[Antonucci(1993)]{antonucci93} 
  Antonucci, R.\ 1993, \araa, 31, 473 

\bibitem[Ashby et al.(2009)]{ashby09} 
  Ashby, M.~L.~N., et al.\ 2009, \apj, 701, 428 

\bibitem[Assef et al., 2008]{assef08} 
  Assef, R.J., Kochanek, C.S., Brodwin, M., Brown, M. J. I., Caldwell,
  N., Cool, R. J., Eisenhardt, P., Eisenstein, D., Gonzalez, A. H.,
  Jannuzi, B. T., Jones, C., McKenzie, E., Murray, S. S., Stern, D.\
  2008, \apj, 676, 286

\bibitem[Assef et al.(2010)]{assef09} 
  Assef, R.~J., et al.\ 2010, \apj, 713, 970 

\bibitem[Baldwin et al.(1981)]{baldwin81} 
  Baldwin, J.~A., Phillips, M.~M., \& Terlevich, R.\ 1981, \pasp, 93,
  5

\bibitem[{Bertin \& Arnouts}, 1996]{sextractor96}	
  Bertin, E. \& Arnouts, S.\ 1996, \aaps, 117, 393

\bibitem[Bower et al.(2006)]{bower06} 
  Bower, R.~G., Benson, A.~J., Malbon, R., Helly, J.~C., Frenk, C.~S.,
  Baugh, C.~M., Cole, S., \& Lacey, C.~G.\ 2006, \mnras, 370, 645

\bibitem[Brand et al.(2006)]{brand06} 
  Brand, K., et al.\ 2006, \apj, 641, 140 

\bibitem[Brown et al., 2006]{brown06}
  Brown, M.J.I. et al.\ 2006, \apj, 638, 88

\bibitem[Cole et al.(2001)]{cole01} 
  Cole, S., et al.\ 2001, \mnras, 326, 255 

\bibitem[Cool et al.(in prep.)]{cool10}
  Cool, R.~J.\ 2010, in preparation. 

\bibitem[Cool(2007)]{cool07} 
  Cool, R.~J.\ 2007, \apjs, 169, 21 

\bibitem[Cool et al.(2006)]{cool06} 
  Cool, R.~J., et al.\ 2006, \aj, 132, 823 

\bibitem[Croom et al., 2004]{croom04}
  Croom, S.M., et al.\ 2004, \mnras, 349, 1397

\bibitem[Croom et al.(2009)]{croom09} 
  Croom, S.~M., et al.\ 2009, \mnras, 399, 1755 

\bibitem[Croton et al.(2006)]{croton06} 
  Croton, D.~J., et al.\ 2006, \mnras, 365, 11 

\bibitem[Dai et al.(2009)]{dai08} 
  Dai, X., et al.\ 2009, \apj, 697, 506 

\bibitem[Dey et al., in prep.]{dey05}
  Dey, A. et al.\ 2010, in preparation.

\bibitem[Di Matteo et al.(2005)]{dimatteo05} 
  Di Matteo, T., Springel, V., \& Hernquist, L.\ 2005, \nat, 433, 604 

\bibitem[Donley et al.(2008)]{donley08}
  Donley, J.~L., Rieke, G.~H., P{\'e}rez-Gonz{\'a}lez, P.~G., \&
  Barro, G.\ 2008, \apj, 687, 111

\bibitem[Elston et al., 2006]{flamex06}
  Elston, R.J., Gonzalez, A. H. et al.\ 2006, \apj, 639, 816

\bibitem[Efstathiou et al.(1988)]{efstathiou88} 
  Efstathiou, G., Ellis, R.~S., \& Peterson, B.~A.\ 1988, \mnras, 232,
  431

\bibitem[Fabricant et al., 2005]{fabricant05}
  Fabricant, D., et al.\ 2005, \pasp, 117, 1411

\bibitem[Fan(1999)]{fan99} 
  Fan, X.\ 1999, \aj, 117, 2528 

\bibitem[Fan et al., 2001]{fan01}
  Fan, X., et al.\ 2001, \aj, 121, 54 

\bibitem[Fazio et al.(2004)]{fazio04} 
  Fazio, G.~G., et al.\ 2004, \apjs, 154, 10 

\bibitem[Gorjian et al.(2008)]{gorjian08} 
  Gorjian, V., et al.\ 2008, \apj, 679, 1040 

\bibitem[Hasinger et al.(2005)]{hasinger05}
  Hasinger, G., Miyaji, T., \& Schmidt, M.\ 2005, \aap, 441, 417 

\bibitem[Hopkins et al.(2009)]{hopkins09} 
  Hopkins, P.~F., Hickox, R., Quataert, E., \& Hernquist, L.\ 2009,
  \mnras, 398, 333

\bibitem[Hopkins et al.(2007)]{hopkins07} 
  Hopkins, P.~F., Richards, G.~T., \& Hernquist, L.\ 2007, \apj, 654,
  731

\bibitem[Hopkins et al.(2006)]{hopkins06} 
  Hopkins, P.~F., Hernquist, L., Cox, T.~J., Di Matteo, T., Robertson,
  B., \& Springel, V.\ 2006, \apjs, 163, 1

\bibitem[Hopkins et al.(2005)]{hopkins05} 
  Hopkins, P.~F., Hernquist, L., Cox, T.~J., Di Matteo, T., Martini,
  P., Robertson, B., \& Springel, V.\ 2005, \apj, 630, 705

\bibitem[Hopkins et al.(2005b)]{hopkins05b} 
  Hopkins, P.~F., Hernquist, L., Cox, T.~J., Di Matteo, T., Robertson,
  B., \& Springel, V.\ 2005, \apj, 630, 716

\bibitem[Kewley et al.(2001)]{kewley01} 
  Kewley, L.~J., Dopita, M.~A., Sutherland, R.~S., Heisler, C.~A., \&
  Trevena, J.\ 2001, \apj, 556, 121

\bibitem[Jannuzi \& Dey, 1999]{ndwfs99} 
  Jannuzi, B. T. \& Dey, A.\ 1999, ASP Conference Series, Vol. 191,
  p. 111

\bibitem[Jannuzi et al., in prep.]{jannuzi05}
  Jannuzi, B.T. et al.\ 2010, in preparation.

\bibitem[Jiang et al.(2009)]{jiang09} 
  Jiang, L., et al.\ 2009, \aj, 138, 305 

\bibitem[Kauffmann et al.(2003)]{kauffmann03} 
  Kauffmann, G., et al.\ 2003, \mnras, 346, 1055 

\bibitem[Kenter et al.(2005)]{kenter05} 
  Kenter, A., et al.\ 2005, \apjs, 161, 9 

\bibitem[Kochanek et al., in prep.]{kochanek09}
  Kochanek, C.S. et al.\ in preparation

\bibitem[Lacy et al.(2004)]{lacy04} 
  Lacy, M., et al.\ 2004, \apjs, 154, 166

\bibitem[Lin et al., 1996]{lcrslf}
  Lin, H., Kirshner, R.P., Shectman, S.A., Landy, S.D., Oemler, A.,
  Tucker, D.L. \& Schechter, P. L.\ 1996, \apj, 464, 60

\bibitem[Martin et al., 2005]{martin05}
  Martin, D.C. et al.\ 2005, \apj, 619L, 1

\bibitem[Miyaji et al.(2001)]{miyaji01} 
  Miyaji, T., Hasinger, G., \& Schmidt, M.\ 2001, \aap, 369, 49

\bibitem[Morrissey et al.(2007)]{morrissey07}
  Morrissey, P., et al.\ 2007, \apjs, 173, 682 

\bibitem[Moustakas et al.(2010)]{moustakas09}
  Moustakas et al.\ 2009, in prep.

\bibitem[Murray et al., 2005]{xbootes}
  Murray, S.S. et al.\ 2005, \apjs, 161, 1

\bibitem[Page \& Carrera(2000)]{page2000} 
  Page, M.~J., \& Carrera, F.~J.\ 2000, \mnras, 311, 433 

\bibitem[Richards et al.(2009)]{richards09} 
  Richards, G.~T., et al.\ 2009, \aj, 137, 3884

\bibitem[Richards et al., 2006a]{richards06a}
  Richards, G.T. et al.\ 2006a, \apjs, 166, 470

\bibitem[Richards et al.(2006b)]{richards06b} 
  Richards, G.~T., et al.\ 2006b, \aj, 131, 2766 

\bibitem[Richards et al., 2005]{richards05}
  Richards, G.T. et al.\ 2005, \mnras, 360, 839

\bibitem[Rieke et al.(2004)]{rieke04} 
  Rieke, G.~H., et al.\ 2004, \apjs, 154, 25 

\bibitem[Rowan-Robinson et al.(2008)]{rowan08} 
  Rowan-Robinson, M., et al.\ 2008, \mnras, 386, 697 

\bibitem[Rujopakarn et al.(2010)]{rujopakarn10}
  Rujopakarn, W., et al.\ 2010, \apj, 718, 1171 

\bibitem[Salvato et al.(2009)]{salvato09} 
  Salvato, M., et al.\ 2009, \apj, 690, 1250 

\bibitem[Sandage et al.(1979)]{sandage79} 
  Sandage, A., Tammann, G.~A., \& Yahil, A.\ 1979, \apj, 232, 352

\bibitem[Schmidt, 1968]{schmidt68}
  Schmidt, M.\ 1968, \apj, 151, 393

\bibitem[Smith et al.(2005)]{smith05} 
  Smith, R.~J., Croom, S.~M., Boyle, B.~J., Shanks, T., Miller, L., \&
  Loaring, N.~S.\ 2005, \mnras, 359, 57

\bibitem[Stern et al.(2007)]{stern07} 
  Stern, D., et al.\ 2007, \apj, 663, 677 

\bibitem[Stern et al., 2005]{stern05}
  Stern, D. et al.\ 2005, \apj, 631, 163

\bibitem[Strateva et al.(2005)]{strateva05} 
  Strateva, I.~V., Brandt, W.~N., Schneider, D.~P., Vanden Berk,
  D.~G., \& Vignali, C.\ 2005, \aj, 130, 387

\bibitem[York et al., 2000]{sdss}
  York D. et al.\ 2000, \aj, 120, 1579

\bibitem[Weedman et al., 2006]{weedman06}
  Weedman, D.W. et al.\ 2006, \apj, 651, 101

\bibitem[Wolf et al.(2003)]{wolf03} 
  Wolf, C., Wisotzki, L., Borch, A., Dye, S., Kleinheinrich, M., \&
  Meisenheimer, K.\ 2003, \aap, 408, 499

\end{thebibliography}
\end{document}